\documentclass[aps,prd,amsfonts,nofootinbib,supercriptaddress]{revtex4}
\usepackage{amssymb,amsmath}
\usepackage{epsfig}

\newcommand{\lbfig}[1]{\refstepcounter{fig} \label{#1} }
\newcounter{fig}

\newcommand{\beq}{\begin{equation}}
\newcommand{\eeq}{\end{equation}}
\newcommand{\bea}{\begin{eqnarray}}
\newcommand{\eea}{\end{eqnarray}}

\newcommand{\sign}{\;\mbox{sign}}

\begin{document}

\hfill Preprint numbers: ITP-UU-11/04, SPIN-11/02


\title{Time transients
       in the quantum corrected Newtonian potential
       induced by a massless nonminimally coupled scalar field}

\author{Anja Marunovic$^{a*}$}
\author{Tomislav Prokopec$^{b}$}
\email[]{Anja.Marunovic@fer.hr, t.prokopec@uu.nl}
\affiliation{$^a$Department of Physics, FER, University
of Zagreb, Unska 3, HR-10 000 Zagreb, Croatia\\
 $^b$Institute for Theoretical Physics and Spinoza
Institute, Utrecht University, Leuvenlaan 4, 3584 CE Utrecht, The
Netherlands
}

\begin{abstract} \noindent

 We calculate the one loop graviton vacuum polarization induced
by a massless, nonminimally coupled scalar field on Minkowski
background. We make use of the Schwinger-Keldysh formalism, which
allows us to study time dependent phenomena.
 As an application we compute the leading
quantum correction to the Newtonian potential of a
point particle. The novel aspect of the
calculation is the use of the Schwinger-Keldysh
formalism, within which we calculate the time transients induced
by switching on the graviton-scalar coupling.

\end{abstract}


\maketitle

\section{Introduction}

 It has been known for a while that general relativity is, when viewed as
a quantum theory, nonrenormalizable. Indeed, when gravity is
coupled to scalar matter, new local counter-terms are required to
cancel the divergences occurring at one
loop~\cite{'tHooft:1974bx}, while in the case of pure gravity, new
counter-terms occur at two
loops~\cite{Goroff:1985th,vandeVen:1991gw}. That does not
necessarily mean that quantizing gravity and studying perturbative
corrections is meaningless, as long as one is interested in low
energy effects, which are much below the Planck scale. This
effective field theory view on quantum gravity has been fruitful,
and has lead to useful
results~\cite{Weinberg:2009bg,Woodard:2009ns,Donoghue:2009mn}.
Observable effects occur primarily in cosmology,
where as a result of tree level
quantization of gravity during inflation, one is led to the
Newtonian potentials~\cite{Mukhanov:1981xt,Mukhanov:1990me} that
leave observable imprint on the relic cosmic microwave radiation,
as well as the seeds for the large scale structure of the
Universe. The question whether perturbative quantum gravitational
effects can change the basic tree level predictions of
inflationary cosmology is still hotly
debated~\cite{PIworkshop:2010,Prokopec:2010be}.

 On the other hand, there are well-known tree level quantum
effects occurring in curved space-times, the most famous one being
the Hawking thermal radiation generated by black holes. There is
even a flat space effect -- the well known example being the Unruh
effect -- which constitutes thermal radiation seen by accelerating
observers in Minkowski space. Neither Hawking nor the Unruh effect
have so far been confirmed experimentally. The quantum corrections
to these tree level effects are as yet not well understood. In
particular, it would be of interest to investigate whether such
quantum corrections can induce a significant back-reaction on the
background space-time that would correct some undesirable
features that occur when fields are quantized on black hole spaces,
such as quasi-normal modes (which essentially tell us that it is
inconsistent to quantize massless fields on black hole
backgrounds). The main message from these studies is that: we
shall be able to incorporate self-consistently the back-reaction
from fluctuating quantum fields on black hole backgrounds only if
we allow space-times to become dynamical. In other words, a
self-consistent semiclassical gravitational theory will require a
(quantum) modification of the Birkhoff theorem (whose classical
version states that a spherically symmetric distribution of matter
can induce static, radially dependent, metrics only).

 In this paper we do not address this very interesting question, but
we point out that the Schwinger-Keldysh
formalism~\cite{Schwinger:1961,Keldysh:1964ud,Jordan:1986,Hu:1987,Weinberg:2005}
is suitable for such studies. As an example of how to apply the formalism,
we provide a perturbative calculation of the quantum correction
to the Newtonian potential generated by a massless scalar field. This
represents a baby version of the more interesting problem
of quantum corrections to black hole space-times.
 Our treatment generalizes the work of Park and Woodard~\cite{Woodard:2010},
in that we consider not just the minimally coupled massless scalar
field, but also include a nonminimal coupling of the scalar field
to the Ricci curvature scalar. Another important difference is in
that Park and Woodard assumed that gravity was turned on at
$t_0\rightarrow -\infty$, thus preventing any time transients from
occurring. We, on the other hand, assume that the scalar field
decouples from gravity at early times ($t<t_0$), and that the
coupling turns on at $t=t_0$, where $t_0$ is some finite time. In
practice, this can be realized, for example, by a Higgs mechanism,
which gives a large mass $m_\phi$ to the scalar at $t<t_0$, and
the mass vanishes at $t\geq t_0$. Such a scalar will effectively
decouple from gravity at $t<t_0$ and on the length scales
$L>1/m_\phi$, and will induce a nontrivial gravitational effect
for $t>t_0$. In this paper we demonstrate that the
Schwinger-Keldysh formalism allows us to follow in a manifestly
causal manner the gravitational field induced by such quantum
scalar field fluctuations.

 Of course, our work is not the first to address quantum corrections
to Newtonian potentials. Such studies were pioneered by
Donoghue~\cite{Donoghue:1994A,Donoghue:1994B},
and followed by many others~\cite{HamberLiu:1995}, \cite{Akhundov:1997},
with differing results. The problem was again revived in the 2000s by
Bjerrum-Bohr \emph{et al.}~\cite{BjerrumBohr:2002B,BjerrumBohr:2003C},
Butt~\cite{Butt:2006}, Faller~\cite{Faller:2008} \emph{etc.}
where also some of the graviton vertex corrections were considered.
Also the works of Dalvit, Mazzitelli, Satz and
Alvarez~\cite{Dalvit:1994gf,Satz:2004hf} are of interest to us. Indeed,
in Appendix~C we show that there is a rather intricate connection
between the methods used in Refs.~\cite{Dalvit:1994gf,Satz:2004hf}
and our methods.

 The paper is organized as follows: in Sec. II we briefly
show how to obtain the classical Newtonian potential. Section III
is devoted to calculation of the one-loop graviton vacuum
polarization induced by a nonminimally coupled scalar field, while
Sec. IV presents an application of the main result of Sec.
III: the quantum one-loop correction to the Newtonian potential.
In Sec. V we summarize our main results.
In the Appendixes we review the Schwinger-Keldysh formalism
(Appendix~A); some subtleties in deriving the Newtonian potential
of a point particle (Appendix~B); we show how to expand the 2PI
scalar bubble diagram to get the stress-energy tensor and the
graviton vacuum polarization tensor (Appendix~C); and we
present details of the one-loop vacuum polarization calculation (Appendix~D).

Unless stated explicitly, we work in natural units where
$\hbar=1=c$.

\section{Classical Newtonian potential}
\label{Classical Newtonian potential}
We are interested in the gravitational response of a static, point-like
particle of a mass $M$ in the particle's rest frame,
with the classical stress energy tensor in Minkowski background
\beq T_{\mu\nu}^{(c)}=M\delta_\mu^0\delta_\nu^0\delta^{D-1}(\vec x\,),
\label{stress energy:point mass}
 \eeq
where $D$ denotes the dimension of space-time.
For gravity we take the classical Einstein-Hilbert action:
\beq S_{EH}=-\frac{1}{\kappa^2}\int d^Dx\sqrt{-g}R \qquad
(\kappa^2=16\pi G_N)\,, \label{Einstein-Hilbert} \eeq
where $R$ is the Ricci scalar, $g$ is the determinant of the metric tensor
$g_{\mu\nu}$, and $G_N$ is the Newton constant.\\
In order to get the classical Newtonian potential, we first expand
the metric tensor around the flat Minkowski metric $\eta_{\mu\nu}$:
\beq
g_{\mu\nu}=\eta_{\mu\nu}+ h_{\mu\nu}
\,,\qquad\qquad
  \eta_{\mu\nu} = {\rm diag} (-1,\underbrace{1,..,1}_\text{D-1}),
\label{gmunu:expansion}
\eeq
where, in general, $h_{\mu\nu}=h_{\mu\nu}(x)$, $x=(t,\vec x)$. The
full classical gravitational response to a point mass is obtained
by varying the action~(\ref{Einstein-Hilbert}) and setting it
equal to the classical stress-energy tensor, $ \delta S_m / \delta
h^{\mu\nu}=-\sqrt{-g}/2 T_{\mu\nu}$. Here we are primarily
interested in the leading gravitational response (Newtonian
potential) to a point particle at rest, whose stress-energy tensor
is given by~(\ref{stress energy:point mass}). To accomplish this
we need the Einstein-Hilbert action to quadratic order in
$h_{\mu\nu}$ (for detailed calculations see Appendix B):

\beq S_{EH}=\frac{1}{2\kappa^2}\int d^Dx
         \Big[h^{\mu\nu}L_{\mu\nu\rho\sigma}h^{\rho\sigma}
           + {\cal O}(h_{\mu\nu}^3)
         \Big],
\label{Einstein-Hilbert:2}
\eeq
where $L_{\mu\nu\rho\sigma}$ stands for the Lichnerowicz
operator in Minkowski background:
\beq
L_{\mu\nu\rho\sigma}
 = \partial_{(\rho}\eta_{\sigma)(\mu}\partial_{\nu)}
  - \frac{1}{2}\eta_{\mu(\rho}\eta_{\sigma)\nu}\partial^2
  - \frac{1}{2}(\eta_{\mu\nu}\partial_{\rho}\partial_\sigma
               +\eta_{\rho\sigma}\partial_\mu\partial_\nu)
  +\frac12\eta_{\mu\nu}\eta_{\rho\sigma}\partial^2
\,.
\label{Lichnerowicz}
\eeq
 For a later use we rewrite the Lichnerowicz operator
as a combination of two simpler operators:
\beq
L_{\mu\nu\rho\sigma} = D_{\mu\nu\rho\sigma}
                     - \frac{1}{2}\eta_{\mu\nu}D_{\rho\sigma}
\,,
\eeq
where
\beq
D_{\rho\sigma}=\partial_{\rho}\partial_\sigma-\eta_{\rho\sigma}\partial^2
\;,\qquad D_{\mu\nu\rho\sigma} =
\partial_{(\rho}\eta_{\sigma)(\mu}\partial_{\nu)}
         - \frac{1}{2}\eta_{\mu(\rho}\eta_{\sigma)\nu}\partial^2
         - \frac{1}{2}\eta_{\rho\sigma}\partial_\mu\partial_\nu
\,.
\label{D1:D2}
\eeq
The Lichnerowicz operator~(\ref{Lichnerowicz}) possesses
the following symmetries: it is symmetric under the exchange of
the first two indices ($\mu\leftrightarrow \nu$); of the last two indices
($\rho\leftrightarrow \sigma$), as well as under the exchange of
the first two and the last two indices
($\mu\nu\leftrightarrow \rho\sigma$).

 Now varying the action~(\ref{Einstein-Hilbert:2}) with
the stress energy tensor~(\ref{stress energy:point mass}) yields
\beq
L_{\mu\nu\rho\sigma}h^{\rho\sigma}(x)=\;\frac{\kappa^2}{2}\;\delta_\mu^0\delta_\nu^0
M\delta^{D-1}(\vec x).
 \label{EOM cl}
 \eeq
The same equation can be obtained by linearizing the Einstein equation
around the Minkowski background, which is a simpler procedure.
 As we show in Appendix~B, the Newtonian potential in
the longitudinal (Newton) gauge Eq.~(\ref{EOM cl}) gives
the solution:
\beq h_{00}^{(0)}(x)=h_{ii}^{(0)}(x)=\frac{2G_N M}{r},\qquad
(i=1,2,3)
\,,
\label{classical solution}
\eeq
where the superscript $(0)$ emphasizes that the metric components
$h_{\mu\nu}^{(0)}$ refer to the classical solution.
In the following two sections we show how to use this solution
to construct the quantum corrected one-loop Newtonian potential.

\section{Graviton vacuum polarization}
\label{Graviton vacuum polarization}
 Here we derive the leading quantum
contribution to the graviton vacuum polarization tensor due to a
nonminimally coupled massless scalar field $\varphi$, with the
action:
\begin{equation}
S_\varphi=\int d^D x\sqrt{-g}\left\{-\frac{1}{2}(\partial_\mu \varphi)(\partial_\nu \varphi)g^{\mu\nu}-\frac{1}{2}\xi R\varphi^2\right\},
\label{ActionScalar}
\end{equation}
where $\xi$ measures the coupling strength
of $\varphi$ to gravity through the Ricci scalar $R$.
 In this paper we focus on the quantum corrections around Minkowski space,
hence we expand the metric tensor
around the flat Minkowski metric ({\it cf.}
Eq.~(\ref{gmunu:expansion})):
\beq
g_{\mu\nu}=\eta_{\mu\nu}+ \kappa h_{\mu\nu}
\,.
\label{metric}
\eeq
Since we are here primarily interested in the one-loop graviton
vacuum-polarization, in~(\ref{metric}) we have extracted the
gravitational coupling constant $\kappa$ defined in
Eq.~(\ref{Einstein-Hilbert}), as $\kappa^2$ can be used as the
loop counting parameter of perturbative gravity.
 Note that $h_{\mu\nu}(x)$ in Eq.~(\ref{metric}) can be understood
to contain both classical and quantum contributions
to the Newtonian potential, which is in the spirit of the rest of the paper.
 One can write Eq.~(\ref{metric}) in the more general form,
$g_{\mu\nu} = g^{(b)}_{\mu\nu} +\kappa h_{\mu\nu}$, where
$g^{(b)}_{\mu\nu}$ denotes a classical background metric
around which one expands.
In the context of this work, the natural choices for
$g^{(b)}_{\mu\nu}$ are the metric of the classical Newtonian potential,
$g^{(b)}_{\mu\nu}=\eta_{\mu\nu}+h_{\mu\nu}^{(0)}$ with the elements of
$h_{\mu\nu}^{(0)}$ given in Eq.~(\ref{classical solution}), or
by the Schwarzschild black hole metric.
Since choosing these background metrics would entail significant technical
complications, we leave their treatment for a future work.
\\
In our model~(\ref{Einstein-Hilbert}) and~(\ref{ActionScalar}) the
graviton acquires quantum contributions both from the graviton and
scalar quantum fluctuations. To calculate the graviton one loop
vacuum polarization one needs the cubic and quartic graviton
vertices, which can be quite straightforwardly extracted from
Eq.~(\ref{expanding Ricci}). For simplicity in this paper we focus
on the one-loop scalar contribution to the graviton self-energy
(vacuum polarization). To this end we need the cubic and quartic
scalar-graviton vertices, which can be easily extracted from the
scalar action~(\ref{ActionScalar}):
\bea
 S_{h\varphi\varphi}&=&-\frac{\kappa}{2}\int d^D x
\left\{\left(\frac{1}{2}\eta^{\mu\nu}\eta_{\rho\sigma}
-\delta_{(\rho}^{(\mu}\delta_{\sigma)}^{\nu)}\right)(\partial_\mu\varphi(x))(\partial_\nu\varphi(x))+\xi
(D_{\rho\sigma}\varphi^2(x)) \right\}h^{\rho\sigma}(x)
\label{Scubic}
\\
S_{h^2\varphi^2}&=&-\frac{\kappa^2}{2}\int d^D x\;
\bigg\{h^{\mu\nu}(x)\bigg[
\left(\frac{1}{8}\eta^{\alpha\beta}\eta_{\mu\nu}\eta_{\rho\sigma}
-\frac14\eta_{\mu\nu}\delta^\alpha_{(\rho}\delta^\beta_{\sigma)}
-\frac14\eta_{\rho\sigma}\delta^\alpha_{(\mu}\delta^\beta_{\nu)}
-\frac{1}{4}\eta^{\alpha\beta}\eta_{\mu(\rho}\eta_{\sigma)\nu}
+\delta^\alpha_{(\mu}\eta_{\nu)(\rho}\delta_{\sigma)}^\beta
\right)
\nonumber\\
&&\hskip 3.5cm
\times\,(\partial_\alpha\varphi(x))(\partial_\beta\varphi(x))
\bigg]\;h^{\rho\sigma}(x)
\label{Squartic}
\\
\nonumber
&&\qquad\qquad\qquad
+\,\xi\varphi^2(x)
    \bigg[\Big(\frac{h}{2}\Big)
                       (\partial_\mu\partial_\nu h^{\mu\nu}-\partial^2 h)
           - h_{\rho\sigma}(2\partial^\rho\partial^\mu h^\sigma_{\;\mu}
                 -\partial^2 h^{\rho\sigma})
           + h^{\mu\nu}\partial_\mu\partial_\nu h
\nonumber\\
      &&\qquad\qquad\qquad\qquad\qquad
    -\,\frac32(\partial_\rho h^{\rho\sigma})(\partial^\mu h_{\mu\sigma})
         -\frac14(\partial_\mu h)(\partial^\mu h)
        +(\partial^\mu h)(\partial^\nu h_{\mu\nu})
        +\frac34 (\partial^\mu h^{\nu\sigma})(\partial_\mu h_{\nu\sigma})
\bigg]
\bigg\}
\,,
\nonumber
\eea
where we made use of Eqs.~(\ref{metric:expansion in h}),
(\ref{Ricci scalar:spatial:quadratic})
and~(\ref{Lichnerowicz}--\ref{D1:D2}).

In the spirit of the Schwinger-Keldysh formalism, cubic and quartic
parts contribute to the interaction action as:
\beq
 S_{\rm int}[\varphi^+, h_{\alpha\beta}^+,\varphi^-, h_{\alpha\beta}^-]
 = S_{h\varphi\varphi}[\varphi^+,h_{\alpha\beta}^+]
  - S_{h\varphi\varphi}[\varphi^-, h_{\alpha\beta}^-]
  + S_{h^2\varphi^2}[\varphi^+, h_{\alpha\beta}^+]
  - S_{h^2\varphi^2}[\varphi^-, h_{\alpha\beta}^-]
\,.
\label{Sint:SK}
 \eeq
These interactions can be used to generate the graviton tadpole
and the graviton vacuum polarization induced by the one-loop
scalar fluctuations. The scalar diagram that contributes to the
graviton tadpole is shown in Fig.~\ref{figure 1}.
\begin{figure}[ht]
\centerline{\epsfig{file=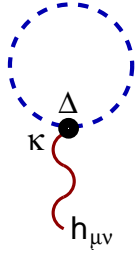,width=1.7cm}}
\lbfig{figure 1} \vskip -0.4cm \caption{The graviton one-loop
tadpole. Only the scalar loop contribution is shown. The scalar
propagator $\Delta$ is the blue dashed line, while the graviton
insertion $h_{\mu\nu}$ is the red solid wavy line, and
$\kappa=\sqrt{16\pi G_N}$ is the gravitational coupling constant.}
\end{figure}

The three scalar field diagrams that
contribute to the  one-particle irreducible (1PI)
 graviton vacuum polarization at the
one-loop order are shown in Fig.~\ref{figure 2}.
\begin{figure}[ht]
\centerline{\epsfig{file=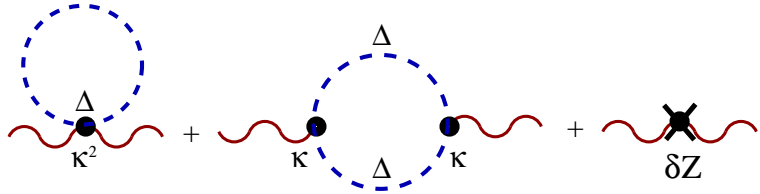,width=10cm}}
\lbfig{figure 2} \vskip -0.2cm \caption{The scalar diagrams that
contribute to the graviton one-loop vacuum polarization. $\delta
Z$ stands for the one-loop counter-terms.}
\end{figure}
The first diagram is the
local contribution generated by the quartic
action~(\ref{Squartic}), the second is the non-local contribution
generated by the cubic action~(\ref{Scubic}) squared, and the third is the
counter-term.

 In this paper we work with the 1PI effective action, which consists
of the tree-level (Einstein-Hilbert) part, the matter free action, and
the one-loop contributions with zero, one, two, three, etc. graviton
insertions. The one loop bubble diagrams with no graviton insertions
are shown in Fig.~\ref{figure 3}.
\begin{figure}[ht]
\centerline{\epsfig{file=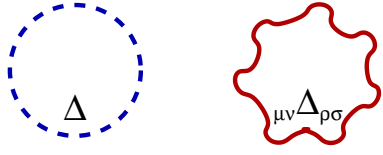,width=5cm}}
\lbfig{figure 3}
\vskip -0.2cm
\caption{The scalar and graviton one-loop bubble diagrams.
${}_{\rho\sigma}\Delta_{\mu\nu}$ denotes the graviton propagator.}
\end{figure}
 Formally, we can write the one-loop 1PI contributions to the effective action
as,
\bea
 \Gamma_1[h_{\mu\nu},\Delta] &=&
          -\frac{\imath}{2}{\rm Tr}\ln[\Delta^{aa}(x;x)]
          -\frac{\imath}{2}{\rm Tr}
                \ln[\phantom{!}_{\mu\nu}\Delta^{aa}_{\rho\sigma}(x;x)]
-\frac12\sum_{a,b=\pm}\int d^Dx
  h_a^{\mu\nu}(x)a\delta^{ab} T^{(1)ab}_{\mu\nu}(x)
\nonumber\\
   &-&\frac12\sum_{a,b=\pm}\int d^D x\int d^D x^\prime
            h_a^{\mu\nu}(x)
   \left[\!\phantom{!}_{\mu\nu}^{\;\;a}\Pi_{\rho\sigma}^b\right](x;x^\prime)
  h_b^{\rho\sigma}(x^\prime)
\label{1loopGamma:SK}
\\
   &&\hskip -1.4cm
-\,\sum_{n=3}^\infty\frac{1}{n!}\!\!\sum_{a_1,a_2,..,a_n=\pm}
      \int d^D x_1 h^{\mu_1\nu_1}_{a_1}(x_1)
       \int d^D x_2 h^{\mu_2\nu_2}_{a_2}(x_2)\cdots\!\!
        \int d^D x_n h^{\mu_n\nu_n}_{a_n}(x_n)
V_{\mu_1\nu_1\mu_2\nu_2..\mu_n\nu_n}^{a_1a_2\cdots a_n}(x_1;x_2;\dots;x_n)
\,,
\nonumber
\eea
where $\Delta^{ab}(x;x^\prime)$ denotes the scalar
Schwinger-Keldysh propagator, which for a massless scalar field on
Minkowski space can be found in Eqs.~(\ref{massless scalar
propagators}--\ref{distance functions}) in Appendix~A, ${\rm Tr}$
denotes both a trace over the space-time indices and an
integration over space-time, $T^{(1)aa}_{\mu\nu}(x)$ denotes the
tadpole in Fig.~\ref{figure 1},
$\left[\phantom{!}_{\mu\nu}^{\pm}\Pi_{\rho\sigma}^\pm\right](x;x^\prime)$
is the graviton vacuum polarization tensor, for which the
contributing one-loop diagrams are shown in Fig.~\ref{figure 2},
and $V_{\mu_1\nu_1\mu_2\nu_2..\mu_n\nu_n}^{a_1a_2\cdots
a_n}(x_1;x_2;\dots;x_n)$ stand for the one-loop vertex corrections
with $n\geq 3$ graviton insertions. An example of a vertex
correction diagram is shown in Fig.~\ref{figure 4}.
\begin{figure}[ht]
\centerline{\epsfig{file=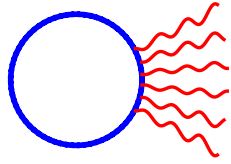,width=3cm}}
\lbfig{figure 4}
\vskip -0.2cm
\caption{A one-loop vertex diagram with one scalar loop
and six graviton insertions.}
\end{figure}
In order to get the complete one-loop correction to the graviton,
one needs to evaluate all of the diagrams contributing to the
one-loop effective action~(\ref{1loopGamma:SK}). This is hard.
However, the corrections with more and more graviton insertions
are expected to be progressively smaller and smaller, and hence it
is often enough to calculate the leading non-vanishing correction.
In this paper we evaluate this leading order correction, which
corresponds to two-graviton insertions. In fact, all leading order
quantum effects on the graviton are embodied in the graviton
vacuum polarization tensor, and the contributing diagrams are
shown in Fig.~\ref{figure 2}.

 The effective action~(\ref{1loopGamma:SK}) can be derived by
a Legendre transform of the free action. Here we sketch a heuristic derivation
of~(\ref{1loopGamma:SK}), which misses the scalar and graviton bubble diagrams.
 The loop contributions to the effective action can loosely be defined
as an expectation value of ${\rm e}^{\imath S_{\rm int}}$, where
$S_{\rm int}$ is the interaction action given in~(\ref{Sint:SK}):
\begin{equation}
 {\rm e}^{\imath \Gamma_1} = \langle{\rm e}^{\imath S_{\rm int}}\rangle
  = 1 + \imath \langle S_{\rm h\varphi\varphi}^+ - S_{\rm h\varphi\varphi}^-\rangle
      + \imath \langle S_{\rm h^2\varphi^2}^+ - S_{\rm h^2\varphi^2}^-\rangle
      -\frac 12 \langle (S_{\rm h\varphi\varphi}^+ - S_{\rm h\varphi\varphi}^-)^2\rangle
    +{\cal O}\Big((h_{\mu\nu}^\pm)^3\Big)
\,, \label{effective action A}
\end{equation}
where $\langle\cdot\rangle$ denotes an expectation value with
respect to a chosen state $|\Omega\rangle$,
 which in our case is the Minkowski
vacuum, corrected for the fact that the interaction
with the graviton switches on at a finite time $t_0$.
This difference allows us to investigate
the thereby induced time transients, and in particular how causality
affects and limits the growth in time of the terms induced by quantum effects.
 In Eq.~(\ref{effective action A}) we introduced the notation,
$S_{\rm h\varphi\varphi}^\pm=S_{\rm
h\varphi\varphi}[\varphi^\pm,h_{\mu\nu}^\pm]$ and $S_{\rm
h^2\varphi^2}^\pm=S_{\rm
h^2\varphi^2}[\varphi^\pm,h_{\mu\nu}^\pm]$.
 In this work we neglect the vertex corrections
to $\Gamma_1$, which occur at the order
${\cal O}(h_{\mu\nu}^3)$ and higher.
It would be of interest to investigate whether such corrections change any of
the results presented in this work. For a discussion of the role
of vertex corrections to the Newtonian potential see
Refs.~\cite{BjerrumBohr:2002A,BjerrumBohr:2002B}.
The first non-trivial term on the right-hand-side
of~(\ref{effective action A})
yields the one-loop graviton tadpole of Fig.~\ref{figure 1},
and thus the scalar one-loop contribution to the stress-energy.
The second contribution yields
the local one-loop contribution to the graviton self-energy,
shown in the first diagram of Fig.~\ref{figure 2}.
The disconnected part of the third contribution is the tadpole squared,
which is a part of the geometric series that is needed to reconstruct
$\exp(\imath \Gamma_1)$;
while the connected part of the third contribution represents the
nonlocal contribution to the one-loop self-energy,
diagrammatically shown by the middle diagram of Fig.~\ref{figure 2},
{\it etc.}

 More economical is the two-particle irreducible (2PI) formalism,
in which one formally writes the effective action in terms of
scalar propagators $\Delta(x;x^\prime;[g^{(b)}_{\mu\nu}])$ in a
general background $g_{\mu\nu}^{(b)}$. The one-loop contributions
to the 2PI effective action are then the scalar and graviton
bubble diagrams from Fig.~\ref{figure 3},
\beq
 \Gamma^{(1)}_{2PI}[\Delta,\phantom{!}_{\mu\nu}\Delta_{\rho\sigma}]
  =  - \frac{\imath}{2}{\rm Tr}\ln[\Delta^{aa}(y;y;[g_{\mu\nu}^{(b)}])]
          -\frac{\imath}{2}{\rm Tr}
   \ln[\phantom{!}_{\mu\nu}\Delta^{aa}_{\rho\sigma}(y;y;g_{\mu\nu}^{(b)})]
\,,
\label{2pi:1loop}
\eeq
while the two-loop contributions come from the diagrams shown in
Fig.~\ref{figure 5}, {\it etc}.
\begin{figure}[ht]
\centerline{\epsfig{file=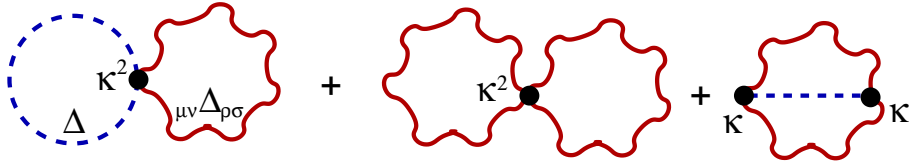,width=12cm}}
\lbfig{figure 5}
\vskip -0.2cm
\caption{The two-loop diagrams contributing to the 2PI effective
action.}
\end{figure}
 For example, Refs.~\cite{Dalvit:1994gf,Satz:2004hf} vary
the first term in Eq.~(\ref{2pi:1loop}) with respect to
$g_{\mu\nu}^{(b)}$, and insert this into the classical linearized
equation for the graviton with a static point mass source to study
the leading quantum correction to the Newtonian potential, which
in that approach originates from the quadratic term in the
expansion of the scalar propagator $\Delta^{ab}$ in
Eq.~(\ref{2pi:1loop}) in powers of $h_{\mu\nu} =
g_{\mu\nu}^{(b)}-\eta_{\mu\nu}$. In Appendix~C we show that that
approach must lead to the same answer for the quantum corrected
Newtonian potential as obtained from the one-loop graviton vacuum
polarization, which is the approach advocated in this paper. A
detailed comparison shows that the results indeed agree,
representing a non-trivial check of our work.

 Let us now go back and discuss various contributions to
Eq.~(\ref{1loopGamma:SK}).
When the scalar and graviton propagators are calculated on Minkowski vacuum,
they do not depend on $h_{\mu\nu}$, and hence
the one-loop bubble terms in~(\ref{1loopGamma:SK})
contribute to the effective action as a constant, and can be neglected.
Therefore, the one-loop graviton effective action in our theory has
the following form ({\it cf.} also Eq.~(\ref{bubble digram:expansion})
in Appendix~C),
\begin{equation}
\Gamma_{1}[h^\pm_{\mu\nu}] = S_{EH}[h_{\mu\nu}^\pm] -\frac12\int
d^Dx h^{\mu\nu}_{\pm}(x)
                             T_{\mu\nu}^{\rm (1)\,\pm}(x)
                        -\frac12\int d^Dx d^Dx^\prime h^{\mu\nu}_\pm(x)
        \big[\phantom{!}_{\mu\nu}^{\pm}\Pi_{\rho\sigma}^\pm\big](x;x^\prime)
                           h^{\rho\sigma}_\pm(x^\prime)
    +{\cal O}\big((h_{\mu\nu}^\pm)^3\big)
\,.
\label{effective action}
\end{equation}
We shall first argue that the tadpole and the local contribution
to the graviton vacuum polarization vanish and then evaluate the
nonlocal contribution to the one-loop graviton vacuum
polarization.

\subsection{Local contributions}
\label{Local contributions}

Since we work in dimensional regularization, the local diagram --
which yields a quadratic divergence -- is automatically
subtracted, and hence will not contribute. To see this in some detail,
from Eq.~(\ref{effective action A}) we infer that
the tadpole contribution to the effective action yields
\beq
\Gamma_{\rm tadpole}=\Gamma_{h\varphi\varphi}[h^{\mu\nu}_\pm]
 =\langle S^+_{h\varphi\varphi}-S^-_{h\varphi\varphi}\rangle
\,,
 \eeq
where $S^\pm_{h\varphi\varphi}$ are defined in Eqs.~(\ref{Scubic})
and~(\ref{Sint:SK}).
If we take look at the structure of $ S^{\pm}_{h\varphi\varphi}$
in Eq.~(\ref{Scubic}) we see that the contributing terms can be written as
\beq
 \langle \varphi(x)^2 \rangle=\imath\Delta_{++}(x;x)
\,\qquad \mbox{and} \qquad
\langle \partial_\alpha\varphi(x)\partial_\beta\varphi(x) \rangle
  =\partial_\alpha\partial_\beta^\prime \imath\Delta_{++}(x;x^\prime)|
           _{x\rightarrow x^\prime}
\,.
\label{tadpole terms zero}
\eeq
 The first term in Eq.~(\ref{tadpole terms zero}) equals
 the coincident ($x=x^\prime$) Feynman propagator
$\imath\Delta_{++}\propto 1/\Delta x_{++}^{D-2}$ (see
Eq.~(\ref{massless scalar propagators})), which is in dimensional
regularization defined as the analytic extension of the propagator
where it is defined, {\rm i.e.} in $\Re[D]<2$, where
$\imath\Delta_{++}(x;x)=0$. Hence,  this term vanishes in the
dimensional regularization. In order to show that the contribution
of the second term vanishes, first note that
$\partial_\alpha\partial_\beta^\prime
\imath\Delta_{++}(x;x^\prime)
 =\delta_\alpha^0\delta_\beta^0\imath\delta^D(x-x^\prime)$. This implies
that when either $\alpha\neq 0$ or $\beta\neq 0$, this term
vanishes, since in dimensional regularization it equals the
analytic extension of $\partial_\alpha\partial_\beta^\prime
\imath\Delta_{++}(x;x^\prime)|
           _{x\rightarrow x^\prime}$ from
$\Re[D]<0$ (where it vanishes) to the whole complex plane. When
both $\alpha=0=\beta$, then this term yields
$\imath\delta^D(x-x^\prime)|_{x^\prime\rightarrow
x}=\imath\delta^D(0)$, which can be shown to vanish in dimensional
regularization. We have thus shown that the tadpole vanishes in
dimensional regularization:
 \beq
 \Gamma_{\rm tadpole}=0
\,.
 \eeq

 Let us next consider the local contribution to the one-loop graviton
vacuum polarization:
 \beq
\Gamma_{\rm local}
   =\Gamma_{h^2\varphi^2}[h^{\mu\nu}_\pm]
    =\langle S^+_{h^2\varphi^2}-S^-_{h^2\varphi^2}\rangle
\,.
 \eeq
 From the structure of the quartic action~(\ref{Squartic})
we see that the same terms as in Eq.~(\ref{tadpole terms zero})
contribute to $\Gamma_{\rm local}$, and therefore, in dimensional
regularization,
 \beq
 \Gamma_{\rm local}=0.
 \eeq
\subsection{Non-local contributions}
As previously elaborated, the only term in Eq.~(\ref{effective action A})
that contributes to
the  one-loop effective action~(\ref{effective action}) is contained
in the Wick-contracted part of the expression
$\imath\delta\Gamma_{h\varphi\varphi}\equiv
 -\frac{1}{2}\big<(S_{h\varphi\varphi}^+-S_{h\varphi\varphi}^-)^2\big>$:
\bea
\delta\Gamma_{h\varphi\varphi}[\varphi^\pm,h_{\alpha\beta}^\pm]
 &=&-\frac{1}{2}\int d^D x\int d^D x^\prime
h^{\mu\nu}_\pm(x)\left[{}^{\pm}_{\mu\nu}\Pi_{\rho\sigma}^{\pm}\right]
   (x;x^\prime)h_\pm^{\rho\sigma}(x^\prime)
\label{GammaCubic}
\\ \nonumber
&=&\frac{\imath}{2}\left(\frac{\pm\kappa}{2}\right)\left(\frac{\pm\kappa}{2}\right)\int
d^D x \,d^D x^\prime h^{\mu\nu}_\pm(x)
\\ \nonumber
&&\times\,\bigg\langle\left[\left(\frac{1}{2}\eta^{\alpha\beta}\eta_{\mu\nu}-\delta_{(\mu}^{(\alpha}\delta_{\nu)}^{\beta)}\right)(\partial_\alpha\varphi_\pm(x))(\partial_\beta\varphi_\pm(x))+\xi (D_{\mu\nu}\varphi^2_\pm(x))\right]
\\ \nonumber
&&\times\,\left[\left(\frac{1}{2}\eta^{\alpha^\prime\beta^\prime}
  \eta_{\rho\sigma}
 -\delta_{(\rho}^{(\alpha^\prime}\delta_{\sigma)}^{\beta^\prime)}\right)
  (\partial^\prime_{\alpha^\prime}\varphi_\pm(x^\prime))
  (\partial^\prime_{\beta^\prime}\varphi_\pm(x^\prime))
  +\xi (D^\prime_{\alpha\beta}\varphi^2_\pm(x^\prime))\right]\bigg\rangle
  h_\pm^{\rho\sigma}(x^\prime)
\,,
\eea
where the primed partial derivatives stand for derivatives with
respect to $x^\prime$.
Next, we perform a detailed calculation of the vacuum polarization
tensor in Appendix D and note that the derivative operators
appearing in the vacuum polarization tensor~(\ref{Self-En-F1}) can
be written in the useful form:
\bea
D_{\alpha\beta\mu\nu}D^{\alpha\beta}{}_{\rho\sigma}
 &=& L_{\mu\nu\alpha\beta}\left(D^{\alpha\beta}{}_{\rho\sigma}
  -\frac{1}{2}\eta^{\alpha\beta}D_{\rho\sigma}\right)
\\
D_{\mu\nu}D_{\rho\sigma}
 &=&L_{\mu\nu\alpha\beta}\left(-\eta^{\alpha\beta}D_{\rho\sigma}\right)
\,,
\eea
where $L_{\mu\nu\alpha\beta}$ is the Lichnerowicz
operator~(\ref{Lichnerowicz}--\ref{D1:D2}).
Making use of the results from Appendix D, we can split the
graviton polarization tensor~(\ref{Self-En-F1}) into a non-local
finite part and a local divergent part as follows:
\beq
 \imath\left[{}^{\pm}_{\mu\nu}\Pi_{\rho\sigma}^{\pm}\right](x;x^\prime)
 = \imath\left[{}^{\pm}_{\mu\nu}\Pi_{\rho\sigma}^{\pm}\right]^{\rm (ren)}
        (x;x^\prime)
 + \imath\left[{}^{\pm}_{\mu\nu}\Pi_{\rho\sigma}^{\pm}\right]^{\rm (div)}
    (x;x^\prime)
\label{Sigma:split:ren+div}
\eeq
where the renormalized  graviton one-loop vacuum polarization
tensor reads
\bea
&&\imath\left[{}^{\pm}_{\mu\nu}\Pi_{\rho\sigma}^{\pm}\right]^{\rm (ren)}
      (x;x^\prime)=-(\pm)(\pm)\frac{\kappa^2}{30720\pi^4}
 \nonumber\\ &&\qquad\qquad\qquad
\times\, L_{\mu\nu\alpha\beta}
   \Bigg\{D^{\alpha\beta}{}_{\rho\sigma}
         -\eta^{\alpha\beta}D_{\rho\sigma}
                 \left[\frac{1}{6}+30\left(\xi-\frac{1}{6}\right)^2\right]
   \Bigg\}
  \partial^4\left[\ln^2(\mu^2\Delta x_{\pm\pm}^2)
                 -2\ln(\mu^2\Delta x_{\pm\pm}^2)\right]
\,.
\label{Self-En-F2}
\eea
This is the generalization of the graviton vacuum polarization
induced by a minimally coupled scalar ($\xi=0$) calculated by Park
and Woodard in~\cite{Woodard:2010} to the case of a massless,
nonminimally coupled, scalar ($\xi\neq 0$), and it is one of the
main results of this work. When $\xi=0$, our
result~(\ref{Self-En-F2}) is in perfect agreement
with~\cite{Woodard:2010}, representing a nontrivial check of our
work.

 The divergent part of the vacuum polarization tensor~(\ref{Self-En-Big})
can be extracted by inserting the divergent part of
Eq.~(\ref{++ExpandFinal}) into (\ref{Self-En-F1}). The result is
 \bea
 \left[{}^{\pm}_{\mu\nu}\Pi^\pm_{\rho\sigma}\right]^{\rm(div)}
 (x;x^\prime)
 &=&(\sigma^3)^{\pm\pm}
\frac{\kappa^2\Gamma\left(\frac{D}{2}\right)\mu^{D-4}}
     {64\pi^{D/2}(D^2-1)(D-2)(D-3)(D-4)}
\label{Self-En-Dvg}
 \\
&&\times\,\Bigg\{4D_{\alpha\beta\mu\nu}D^{\alpha\beta}{}_{\rho\sigma}
+\left[-\frac{D}{D-1}+8(D^2-1)\left(\xi-\frac{D-2}{4(D-1)}\right)^2\right]D_{\mu\nu}D_{\rho\sigma}\Bigg\}
\delta^D(x-x^\prime)
\,.
\nonumber
 \eea
As we show in the next section, this term has the right structure
such that it can be subtracted by local counter-terms, which ought
to be added to the effective action to complete the one-loop
renormalization program.
 \subsection{Renormalization}
 \label{Renormalization}

In order to cancel the divergent part of the graviton vacuum
polarization~(\ref{Self-En-Dvg}), from the form of the effective
action~(\ref{effective action}) we see that it should comprise the
counter-term(s) of the form
\bea
\delta \Gamma_{c.t.}\propto  \int d^Dx d^Dx^\prime h^{\mu\nu}_\pm(x)
   \left[\phantom{!}_{\mu\nu}^{\pm}\Pi_{\rho\sigma}^{\pm}\right]^{\rm(div)}
          (x;x^\prime)h^{\rho\sigma}_\pm(x^\prime)
\,.
\label{Gama c.t.}
\eea
In spite of the appearance, this counter-term~(\ref{Gama c.t.}) is local
(because of the delta function in Eq.~(\ref{Self-En-Dvg})), as it
should be.

As we shall now show, two counter-terms that are quadratic in the
curvature tensors are needed to subtract the divergences
in~(\ref{Self-En-Dvg}). To see this,
first note that the linearized form of Ricci scalar and Ricci tensor
are ({\it cf.} Eq.~(\ref{expanding Ricci}))
 \bea
 R^{\rm (lin)}[h_{\mu\nu}^\pm]
  &=&\partial^\mu\partial^\nu h^\pm_{\mu\nu}-\partial^2 h_\pm
   =D_{\mu\nu}h^{\mu\nu}_\pm
 \label{Ricci scalar lin}
 \\
 R_{\mu\nu}^{\rm (lin)}[h_{\mu\nu}^\pm]
   &=&\Big(\partial_{(\rho}\eta_{\sigma)(\mu}\partial_{\nu)}
           -\frac 12 \eta_{\mu(\rho}\eta_{\sigma)\nu}\partial^2
           -\frac 12\eta_{\rho\sigma}\partial_\mu\partial_\nu  \Big)
               h^{\rho\sigma}_\pm
          = D_{\mu\nu\rho\sigma}h^{\rho\sigma}_\pm
\,,
 \label{Ricci tensor lin}
 \eea
where we expanded the metric tensor around Minkowski space,
$g_{\mu\nu}=\eta_{\mu\nu}+\kappa h_{\mu\nu}$. To quadratic order in
$h_{\mu\nu}$, the following two counter-terms>
\beq
 \delta\Gamma_{\rm c.t.} = \alpha_1\sum_\pm(\pm)
                   \int d^D\sqrt{-g} x R[g^\pm_{\alpha\beta}]^2
    + \alpha_2\sum_\pm(\pm)\int d^D x \sqrt{-g}R_{\mu\nu}[g^\pm_{\alpha\beta}]
                                 R^{\mu\nu}[g^\pm_{\alpha\beta}]
\eeq
and can be recast as
 \bea
 \delta \Gamma_{\rm c.t.} = \sum_{\pm\pm}(\sigma^3)^{\pm\pm}\kappa^2
  \int d^D x d^D x^\prime h^{\mu\nu}_\pm(x)
   \Big[\Big(\alpha_1 D_{\mu\nu}D^\prime_{\rho\sigma}
       +\alpha_2 D_{\alpha\beta\mu\nu}
                 {D^\prime}^{\alpha\beta}_{\;\;\rho\sigma}\Big)
                 \delta^D(x-x^\prime)\Big]
        h^{\rho\sigma}_\pm(x^\prime)
\,.
 \label{Counter terms}
 \eea
Comparing this with Eqs.~(\ref{Self-En-Dvg}) and~(\ref{Self-En-Dvg}),
we can read off the coefficients
$\alpha_1$ and $\alpha_2$ in~(\ref{Counter terms}):
 \bea
\alpha_1 &=&
\frac{\Gamma\left(\frac{D}{2}\right)\mu^{D-4}}
     {128\pi^{D/2}(D^2-1)(D-2)(D-3)(D-4)}
\left[-\frac{D}{D-1}+8(D^2-1)\left(\xi-\frac{D-2}{4(D-1)}\right)^2\right]
+\alpha_1^{\rm (fin)}
\\
\alpha_2 &=&
\frac{\Gamma\left(\frac{D}{2}\right)\mu^{D-4}}
     {32\pi^{D/2}(D^2-1)(D-2)(D-3)(D-4)}
+\alpha_2^{\rm (fin)}
\,,
\label{Self-En-Dvg:alphas}
 \eea
where $\alpha_1^{\rm (fin)}$ and $\alpha_2^{\rm (fin)}$
represent the finite parts of the counter-terms, which are to be fixed by
measurements. One often uses the minimal subtraction scheme, in which
one expands the coefficients $\alpha_1$ and $\alpha_2$ around $1/(D-4)$.
In this case their divergent parts are
\beq
\alpha_1^{\rm (div)}= \frac{1}{960\pi^2}
          \Big[-\frac13+30\Big(\xi-\frac16\Big)^2\Big]\frac{\mu^{D-4}}{D-4}
\;,\qquad {\rm and} \qquad \alpha_2^{\rm
(div)}=\frac{1}{960\pi^2}\frac{\mu^{D-4}}{(D-4)}
\label{alphas:div} \eeq
and -- when $\xi=0$ -- agrees with Eqs.~(3.3) and~(3.34) of
Ref.~\cite{'tHooft:1974bx}. \footnote{In fact, Veltman and 't
Hooft get a factor two larger result for $\alpha_1$ and
$\alpha_2$, but they work with a complex scalar field, which can
be decomposed into two real scalar fields, thus explaining the
difference.} This completes our discussion of renormalization.

\subsection{Retarded Self Energy}
\label{Retarded Self Energy}
By varying the action~(\ref{effective action}) with
respect to the fields $h^{\mu\nu}_+$ or $h^{\mu\nu}_-$ and setting
$h^{\mu\nu}_+=h^{\mu\nu}_-=h^{\mu\nu}$:
\beq
 \frac{\delta\Gamma[h^{\mu\nu}_\pm]}{\delta
   h^{\mu\nu}_\pm}\Big |_{h^{\mu\nu}_+=h^{\mu\nu}_-=h^{\mu\nu}}=0
\,,
\eeq
the quantum-corrected equation of motion
for the metric perturbation $h_{\mu\nu}$ is
\begin{equation}
L_{\mu\nu\rho\sigma}h^{\rho\sigma}(x)
+\int d^4 x^\prime\left[{}_{\mu\nu}\Pi_{\rho\sigma}^{\rm ret}\right]
       (x;x^\prime)h^{\rho\sigma}(x^\prime) + {\cal O}((h_{\mu\nu})^2)
=\frac{\kappa^2}{2}\delta_\mu^0\delta_\nu^0 M\delta^3(\vec x)
\,,
\label{Full EOM}
\end{equation}
where
\begin{equation}
\left[{}_{\mu\nu}\Pi_{\rho\sigma}\right]^{\rm ret}(x;x^\prime)
  =\left[{}^{+}_{\mu\nu}\Pi_{\rho\sigma}^{+}\right]^{\rm (ren)}(x;x^\prime)
  +\left[{}^{+}_{\mu\nu}\Pi_{\rho\sigma}^{-}\right]^{\rm (ren)}(x;x^\prime)
\label{retarded vacuum polarisation}
\end{equation}
is the retarded graviton vacuum polarization, which we use in the
next section to compute the quantum-corrected Newtonian potential.

\section{The Quantum Corrected Newtonian potential}
\label{The Quantum Corrected Newtonian potential}

 Rather than attempting to consistently solve Eq.~(\ref{Full EOM}), we shall
solve it perturbatively, {\it i.e.} we make the following
 perturbative {\it Ansatz} for the graviton field:
\beq
h_{\mu\nu}(x)=h_{\mu\nu}^{(0)}(x)+h_{\mu\nu}^{(1)}(x)
\,,
\eeq
where $h_{\mu\nu}^{(0)}$ stands for the classical
solution~(\ref{classical solution})
obtained by solving the classical part~(\ref{EOM cl}) of
the full equation of motion~(\ref{Full EOM}).

 To find the quantum correction $h_{\mu\nu}^{(1)}$
we need to solve the perturbative equation
\begin{equation}
L_{\mu\nu\rho\sigma}h^{\rho\sigma (1)}(x)+\int d^4 x^\prime
\left[{}_{\mu\nu}\Pi_{\rho\sigma}\right]^{\rm ret}(x;x^\prime)h^{\rho\sigma
(0)}(x^\prime)=0
\,,
\label{quantum EOM}
\end{equation}
where $\left[{}_{\mu\nu}\Pi_{\rho\sigma}\right]^{\rm ret}$ is
given in Eqs.~(\ref{Self-En-F2}) and~(\ref{retarded vacuum
polarisation}). \footnote{Had we attempted to find an exact
solution to (\ref{Full EOM}), we would need to solve an integral
equation for $h^{00} = (2G_NM/r)H_0(\kappa r)$ $h^{00} =
\delta^{ij}(2G_NM/r)H_1(\kappa r)$, where $H_0(z)$ and $H_1(z)$
are the sought-for functions, whose asymptotic series (around
$r=\infty$) begins as $H_{1,2}(z)=1+{\cal O}(1/z^2)$. It would be
of interest to find out what are the coefficients of this
asymptotic series and whether it can be resummed, {\it i.e.}
whether it can be analytically extended to the whole complex
$z$-plane. Understanding this function is important, because it
would tell us whether quantum corrections can resolve the $r=0$
singularity of Newtonian gravity. Pursuing this analysis is,
however, beyond the scope of this work.}

 In order to evaluate $\left[{}_{\mu\nu}\Pi_{\rho\sigma}\right]^{\rm ret}$,
it is useful to split the logarithm function in~(\ref{Self-En-F2})
into its real and imaginary parts:
\bea
\ln(\mu^2\Delta x_{++}^2)&=&\ln\left[\mu^2(-\Delta t^2+\Delta
r^2+\imath\epsilon)\right]=\ln|\mu^2(-\Delta t^2+\Delta
r^2)|+\imath\pi\Theta(\Delta t^2-\Delta r^2)\\ \nonumber
\ln(\mu^2\Delta x_{+-}^2)&=&\ln\left[\mu^2(-\Delta t^2+\Delta
r^2-\imath\epsilon\sign(\Delta t))\right]=\ln|\mu^2(-\Delta t^2+\Delta
r^2)|-\imath\pi\Theta(\Delta t^2-\Delta r^2)\sign(\Delta t)
\,,
 \eea
 with the abbreviations $\Delta r=\|\vec x-\vec x^{\,\prime}\|$ and $\Delta
 t=t-t^\prime$. The retarded graviton vacuum polarization is then
\bea
\left[{}_{\mu\nu}\Pi_{\rho\sigma}^{\rm ret}\right](x;x^\prime)
  &=& -\frac{\kappa^2}{7680\pi^3}\times
L_{\mu\nu\alpha\beta}
\bigg\{D^{\alpha\beta}{}_{\rho\sigma}-\eta^{\alpha\beta}D_{\rho\sigma}
  \bigg[\frac{1}{6}+30\Big(\xi-\frac{1}{6}\Big)^2\bigg]\bigg\} \partial^4
\nonumber\\
&&\hskip 1.6cm
\times\,
 \Theta(\Delta t^2-\Delta r^2)\Theta(\Delta
t)\left[\ln| \mu^2(\Delta r^2-\Delta t^2)|-1\right]
\,.
\label{retarded self energy}
\eea
Recalling that $h^{(0)\rho\sigma}(x^\prime)=2G_N M\delta^{\rho\sigma}/r^\prime$
($r^\prime = \|\vec x^{\,\prime}\|$)
for $\rho=\sigma=0,...,3$ and inserting it in Eq.~(\ref{quantum EOM})
together with~(\ref{retarded self energy}) we get
\bea
L_{\mu\nu\rho\sigma}h^{\rho\sigma (1)}(x)
&=&\frac{\kappa^2 G_NM}{3840\pi^3}
  L_{\mu\nu\rho\sigma}\bigg\{D^{\rho\sigma}{}_{\alpha\beta}
            -\eta^{\rho\sigma}D_{\alpha\beta}
                    \left(\frac16+30(\xi-\frac 16)^2\right)
                     \bigg\}\partial^4F^{\alpha\beta}
\,,
\label{finalEOM}
 \eea
where
\begin{equation}
F^{\alpha\beta}\equiv F(r,t)=\int_{t_0}^t dt'\int d^3
x^\prime \Theta(\Delta t-\Delta r)\left[\ln\mu^2(\Delta t^2-\Delta
r^2)-1\right]\frac{1}{r^\prime}
\,.
\end{equation}
Here we are not interested in homogeneous solutions of
the operator $L_{\mu\nu\rho\sigma}$, and thus
$L_{\mu\nu\rho\sigma}$ can be dropped out from both sides of
Eq.~(\ref{finalEOM}), by which~(\ref{finalEOM}) simplifies considerably.

 It is much easier to carry out the integration in~(\ref{finalEOM})
if we change the variable of integration $\vec x^{\,\prime}$ to
$\Delta \vec r = \vec x-\vec x^{\,\prime}$,
\emph{i.e.}, $d^3x^\prime=d^3\Delta r=2\pi(\Delta r)^2d(\Delta r)
\sin\theta d\theta$. The integration over $\theta$ yields
\begin{equation}
\int_0^\pi\sin\theta d\theta\frac{1}{r^\prime}=
\int_{-1}^1 \frac{d(\cos\theta)}{\sqrt{r^2+\Delta r^2 -2r\Delta r\cos(\theta)}}
=\frac{r+\Delta r-|r-\Delta r|}{r\Delta r }
\,,
\end{equation}
where $r=\|\vec x\,\|$ and we choose $\vec r^{\,\prime}=\hat z r^\prime$.
We thus have
\bea
F(r,t)=4\pi\int_{t_0}^{t}dt^\prime \int_0^\infty d(\Delta r)(\Delta r)^2
 \left\{\ln[\mu^2(\Delta t^2-\Delta r^2)]-1\right\}
\bigg\{\frac{1}{\Delta r}\Theta(\Delta r-r)
  +\frac{1}{r}\Theta(r-\Delta r)\bigg\}\Theta(\Delta t-\Delta r)
\,.
\quad
\label{Frt}
\eea
For simplicity, we decompose this integral into two parts,
namely $I_A$ and $I_B$:
\bea F= I_A+I_B\;,\qquad
 I_A&=&4\pi\int_{t_0}^{t}dt^\prime \int_0^\infty d(\Delta r)(\Delta r)
 \left\{\ln[\mu^2(\Delta t^2-\Delta r^2)]-1\right\}
        \Theta(\Delta r-r)\Theta(\Delta t-\Delta r)\\
I_B&=&\frac{4\pi}{r}\int_{t_0}^{t}dt^\prime
         \int_0^\infty d(\Delta r)(\Delta r)^2
  \left\{\ln[\mu^2(\Delta t^2-\Delta r^2)]-1\right\}
             \Theta(r-\Delta r)\Theta(\Delta t-\Delta r)
\,.\qquad
\eea
If we now take $\Delta t_0\equiv t-t_0$ and carefully treat the
$\Theta$-functions, for $I_A$ we get the following result:
\bea
I_A&=&4\pi\int_{0}^{\Delta t_0}d(\Delta t)\int_{r}^{\Delta t} d(\Delta r)
   \Delta r\left\{\ln[\mu^2(\Delta t^2-\Delta r^2)]-1\right\}\Theta(\Delta t-r)
\\ \nonumber
  &=&\frac{2\pi}{9}
         \bigg\{-8\Delta t_0^3+30\Delta t_0 r^2-22 r^3
               -6r^3\ln\bigg(\frac{\Delta t_0+r}{\Delta t_0-r}\bigg)
        +12 r^3\ln (2\mu r)
\\ \nonumber
&&\hskip 0.5cm
  +\,3\Delta t_0(\Delta t_0^2-3r^2)
                \ln\big[\mu^2(\Delta t_0^2-r^2)\big]
         \bigg\} \Theta(\Delta t_0-r)\;.
\eea
In order to calculate $I_B$, we first write it as a sum of two $\Theta$-functions:
\bea
I_B=\frac{4\pi}{r}\int_{0}^{\Delta t_0}d\Delta t
&& \bigg\{ \int_0^r d(\Delta r)\Delta r^2
  \left\{\ln[\mu^2(\Delta t^2-\Delta r^2)]-1\right\}\Theta(\Delta t-r)
\\ \nonumber
&&+\,\int_0^{\Delta t} d(\Delta r)\Delta r^2
  \left\{\ln[\mu^2(\Delta t^2-\Delta r^2)]-1\right\}\Theta(r-\Delta t)\bigg\}
\,.
\eea
Note that, after integrating over $\Delta r$,  the second integral in $I_B$
gets divided in two parts, thus explaining the origin of
the time transients. Schematically this means
\bea
\int_0^{\Delta t_0}d\Delta t\Big\{\quad \cdots\quad\Big\}\Theta(r-\Delta t)
  \longrightarrow
\int_0^{\Delta t_0}d\Delta t\Big\{ \quad \cdots\quad \Big\}\Theta(r-\Delta t_0)
+\int_0^{r}d\Delta t\Big\{ \quad \cdots\quad \Big \}\Theta(\Delta t_0-r)
\,.
\eea
It is now convenient to break the overall result for $F$ into a sum over two parts,
each proportional to one $\Theta$-function:
\begin{equation}
F(r,t)\equiv F_1(r,t)\Theta(\Delta t_0-r)+ F_2(r,t)\Theta(r-\Delta
t_0)\;,
\end{equation}
where $F_1$ and $F_2$ evaluate to:
\begin{eqnarray}
F_1(r,t)&\equiv& 4\pi\frac{r^3}{6}\left[\ln(2\mu r)-\frac{25}{12} \right]
\label{F1}
\\
 &&\hskip -0.3cm +\,4\pi\frac{\Delta
t_0^2-r^2}{6}\left[\frac{(\Delta t_0+r)^2}{2r}\ln[\mu(\Delta t_0+r)]
    -\frac{(\Delta t_0-r)^2}{2r}\ln[\mu(\Delta t_0-r)]
   -\frac{11}{3}\Delta t_0 \right]\\
F_2(r,t)&\equiv& 4\pi\frac{\Delta t_0^4}{6r}
  \left[ \ln(2\mu\Delta t_0)-\frac{25}{12} \right]
\label{F2}
\,.
\end{eqnarray}
One can easily verify that $\partial^2 F$ satisfies
\begin{equation}
\partial^2F(r,t)=\left[\partial^2F_1(r,t)\right]\Theta(\Delta
t_0-r)+\left[\partial^2F_2(r,t)\right]\Theta(r-\Delta t_0)
\,,
\label{partialF}
\end{equation}
and analogously for $\partial^4F$. This simple result holds true
because, not only $F_1$ and $F_2$ are continuous at the causal boundary
$\Delta t_0 = r$, but also the pairs
$\{(\partial_0+\partial_r) F_1,(\partial_0+\partial_r)F_2\}$,
$\{\partial^2 F_1,\partial^2F_2\}$ and
$\{(\partial_0+\partial_r)\partial^2 F_1,
 (\partial_0+\partial_r)\partial^2F_2\}$ are all continuous
at $\Delta t_0 = r$.
Finally, from~(\ref{F1}--\ref{F2}) and~(\ref{partialF}) we obtain:
\begin{equation}
\partial^4 F=16\pi\left \{\frac{\ln(2\mu r)}{r}\Theta(\Delta t_0-r)+\frac{\ln(2\mu\Delta t_0)}{r} \Theta(r-\Delta t_0)  \right\}
\,.
\end{equation}
By inserting this into~(\ref{finalEOM}), we get the desired result for
the quantum one-loop correction to the Newtonian potential:
\begin{eqnarray}
h^{00(1)}&=&\frac{\kappa^2G_N M}{160\pi^2}\left\{
\frac{1}{3r^3}\left(1+2\tilde\xi\right)\Theta(\Delta t_0-r)
-\frac{1}{r \Delta
t_0^2}\left(-1+2\tilde\xi\right)\Theta(r-\Delta
t_0) +\frac 43\frac 1{r\Delta t_0}(1-\tilde\xi)\delta(r-\Delta t_0)\right\}
\quad\;\;
\label{h00-final}
\\
h^{ij(1)}&=&\frac{\kappa^2G_N M}{160\pi^2}\delta^{ij}\left\{
-\frac{1}{3r^3}(-1+2\tilde\xi)\Theta(\Delta t_0-r)
+\frac{1}{3 r\Delta
t_0^2}\left(-1+6\tilde\xi\right)\Theta(r-\Delta t_0)
+\frac{4\tilde\xi}{3 r\Delta t_0}\delta (r-\Delta
t_0)\right\}
\label{hii-final}
\\
h^{i0(1)}&=&\frac{\kappa^2G_N
M}{240\pi^2}\partial^i\left\{\frac{1}{r\Delta t_0}\Theta(r-\Delta t_0)
 \right\}
\,,
\label{hi0-final}
\end{eqnarray}
where $\tilde\xi$ denotes
\beq
\tilde\xi\equiv\frac 16+30\left(\xi-\frac 16\right)^2
\,.
\label{xi tilde}
\eeq
 Eqs.~(\ref{h00-final}--\ref{xi tilde})
constitute our second main result. Apart from the static results,
which are reproduced in the limit when $\Delta t_0\rightarrow
\infty$, Eqs.~(\ref{h00-final}--\ref{hi0-final}) contain time
transients both on the light cone (where $\Delta t_0 = r$) and
outside the lightcone (where $\Delta t_0 < r$). The coincident
time divergences $\propto 1/(\Delta t_0)^2$ and $\propto 1/(\Delta
t_0)$ in~(\ref{h00-final}--\ref{hi0-final}) outside the light cone
signal initial time divergences, which occur as a consequence of
the sudden switching of the graviton-scalar coupling, and can be
removed by either resorting to an adiabatic
switching~\cite{Koksma:2009wa}, or by a suitably modifying the
initial state at $t=t_0$~\cite{Kahya:2009sz,Garny:2009ni}.
However, performing this initial surface renormalization is beyond
the scope of this work. The appearance of the time transients
in~(\ref{h00-final}--\ref{hi0-final}) is the main novelty of our
approach to quantum gravitational corrections on classical space
times, and can be of help to understanding the dynamical quantum
gravitational back-reaction.

 In the limit when $\Delta t_0\rightarrow \infty$ ($t_0\rightarrow -\infty$)
the second and the third part
of $h^{00(1)}$ and $h^{ij(1)}$ in~(\ref{h00-final}--\ref{hii-final})
vanish and all of $h^{i0(1)}$ in~(\ref{hi0-final}) vanishes, and
one recovers the static result.
 This is to be expected, since in the limit when $t_0\rightarrow -\infty$
time transients should disappear. In this limit our results reduce
to the ones obtained by the {\it in-out} formalism, in which
calculations are usually performed in momentum space, and no time
transients are allowed. Indeed
Refs.~\cite{Dalvit:1994gf,Satz:2004hf} have evaluated the one-loop
correction for a nonminimally coupled scalar in the same
(longitudinal or Newtonian) gauge, albeit by different techniques
(see Appendix~C). Their results are shown in Eqs.~(31)
and~(32-33). When compared with
ours~(\ref{h00-final}--\ref{hii-final}) (with $\Delta
t_0\rightarrow \infty$), the coefficients of $h^{00(1)}$ and
$h^{ij(1)}$ agree in magnitude, but they have the opposite sign.
Comparing, however, with Park and Woodard~\cite{Woodard:2010},
which calculate the effect of a minimally coupled scalar only, our
results agree perfectly. To show that, observe first that in the
longitudinal gauge we use, and in the limit $\Delta t_0\rightarrow
\infty$ ({\it cf.} Eqs.~(\ref{h00-final}--\ref{hii-final})
and~(\ref{App:Bardeen potentials})) we get the quantum-corrected
Bardeen potentials
\beq
 \Phi = -\frac{G_NM}{r}  - \frac{G_N^2M}{60\pi}\frac{1+2\tilde \xi}{r^3}
        +{\cal O}\Big(\frac1{r^5}\Big)
\;,\qquad
 \Psi = -\frac{G_NM}{r}  - \frac{G_N^2M}{60\pi}\frac{1-2\tilde \xi}{r^3}
        +{\cal O}\Big(\frac1{r^5}\Big)
\,,
\label{bardeen potentials:quantum}
\eeq
whose leading quantum parts in the $\xi=0$ ($\tilde\xi=1$) case are,
\beq
 \Phi^{(1)} \;\stackrel{\xi\rightarrow 0}{\longrightarrow}\;
                  -\frac{G_N^2M}{20\pi r^3}
\;,\qquad
 \Psi^{(1)} \;\stackrel{\xi\rightarrow 0}{\longrightarrow}\;
           \frac{G_N^2M}{60\pi r^3}
\,.
\label{bardeen potentials:quantum:xi0}
\eeq
On the other hand, relation~(\ref{ParkWoodard gauge:3})
and Eqs.~(55-56) of Ref.~\cite{Woodard:2010} tell us that,
in the gauge used in Ref.~\cite{Woodard:2010},
${h_{00}}^{(1)}_{\rm PW}=G_N^2M/(10\pi r^3)$,
$h^{(1)}_{\rm PW}=-G_N^2M/(10\pi r^3)$ and
$\tilde h^{(1)}_{\rm PW} = - G_N^2M/(30\pi r)$. Inserting these into
Eqs.~(\ref{App:Bardeen potentials}) and~(\ref{ParkWoodard gauge:4}) we get
$\Phi^{(1)}_{\rm PW} = -G_N^2M/(20\pi r^3)$ and
$\Psi^{(1)}_{\rm PW} = G_N^2M/(60\pi r^3)$, which perfectly agree
with~(\ref{bardeen potentials:quantum:xi0}).

 An interesting question is:
in which physical situations one expects to see time transients
of the kind exhibited in Eqs.~(\ref{h00-final}--\ref{hi0-final}).
One such situation would be
a very massive scalar field which at time $t=t_0$ suddenly becomes
massless ({\it e.g.} via a Higgs-like mechanism).
Such a scalar would effectively decouple from gravity at $t<t_0$,
and would begin acting gravitationally
at $t =t_0$. Equations~(\ref{h00-final}--\ref{hi0-final})
describe a good approximation to the gravitational
field generated by such a scalar.
If the Higgs mechanism would induce a significant change in
the vacuum energy, one would need to consider the change in the Newtonian potential
from a de Sitter background to a Minkowski background. While certainly interesting, this calculation would be technically demanding. For a recent calculation
of the one-loop vacuum polarization in de Sitter background see Ref.~\cite{Park:2011ww}.
One may object that the work presented here is mainly of academic interest.
While this is to an extent true, we emphasize that there are many similar situations
that are realized in nature, where time transients are of crucial importance.
One such example is a shell of collapsing
matter, that eventually forms a black hole.
While understanding the quantum back-reaction in such
a dynamical setting is beyond the scope of the (momentum space)
{\it in-out} formalism, as the present work
indicates, it is well within the scope of an {\it in-in} treatment.

\section{Conclusions}
\label{Conclusions}

 We have calculated the one-loop graviton vacuum polarization
 induced by a massless non-minimally coupled scalar field~(\ref{Self-En-F2})
by making use of the Schwinger-Keldysh formalism.
We have then applied our result to calculate the perturbative quantum correction
to the Newtonian potential of a point particle. The novelty of our approach are
the time transients~(\ref{h00-final}--\ref{xi tilde}),
which naturally appear within the Schwinger-Keldysh formalism. Such transients
would occur, for example, in the case when the scalar mass would change from
a large value to zero by a Higgs-like mechanism.

 There are many directions in which the calculation presented in this paper
might be developed. One such direction is to consider quantum effects due to fermionic, vector and tensor fields. Other possible avenues comprise a study of time transients induced by quantum effects in the process of formation of compact stars, such as neutron stars, black holes, boson stars or gravastars~\cite{Horvat:2011ar}. In particular, these types of studies would improve our understanding of the question of quantum backreaction. For example, growing time transients would indicate that the background space-time ought to be modified.

\section*{Appendix A: Schwinger-Keldysh formalism}
\label{Appendix A: Schwinger-Keldysh formalism}

In this section we give a short overview of the Schwinger-Keldysh
or \emph{in-in}
formalism~\cite{Schwinger:1961,Keldysh:1964ud,Jordan:1986,Hu:1987,Weinberg:2005} by making a comparison with the more conventional
\emph{in-out} formulation of quantum field theory (QFT). Here we
shall follow mainly Refs.~\cite{Jordan:1986,Prokopec:2010be}.
Within the {\it in-out} method the vacuum-to-vacuum amplitudes
$\langle\Omega,t_{\rm out}|\Omega,t_{\rm in}\rangle_J$ are calculated in
the presence of an external source $J$ in virtue of
the path-integral representation of the in-out generating functional $Z[J]$:
\beq
\langle\Omega,t_{\rm out}|\Omega,t_{\rm in}\rangle_J=Z[J]
 =\int \mathcal D\varphi\exp\bigg[\imath\Big(S[\varphi]+\int d^DxJ\varphi
                            \Big)\bigg]
\,,
\label{in-out transition amplitude}
\eeq
where $|\Omega,t_{\rm in/out}\rangle$ represent the {\it in} and
{\it out} states in the remote past and future, respectively,
$S[\varphi]$ is the free (classical) action for a (scalar) field $\varphi(x)$,
$J(x)$ is a source current, and $\mathcal D\varphi=\prod_x d\varphi(x)$ is
a path integral integration measure.
In flat space-time the {\it in} and {\it out} states
are mostly expressed as a superposition of free plane waves defined globally;
in curved spaces however the {\it in} and {\it out} states can differ
(in that they are defined only on part of the space)
and the {\it out} states may even not be known.
For these type of situations the {\it in-in} formalism is more suitable.
The partition function $Z[J]$ of the {\it in-out} formalism generates
time-ordered correlation functions between different states:
\beq
\langle\Omega,t_{\rm out}|T\{\varphi(x_1)\cdots\varphi(x_n)\}
 |\Omega,t_{\rm in}\rangle
 =(-\imath)^n\frac{\delta^n}{\delta J(x_1)\cdots\delta J(x_n)} Z[J]|_{J=0}
\,.
\label{in-out expectation value}
\eeq
In nontrivial gravitational backgrounds the final state of a system
usually cannot be predefined, and hence one resorts to calculating
expectation values with respect to an {\it in} state.
In doing so one first defines the {\it in-in} generating functional
by means of the \emph{two} external
sources $J_+$ and $J_-$:
\beq
Z[J_-,J_+]= _{J_-}\langle\Omega,t_{\rm in}|\Omega,t_{\rm in}\rangle_{J_+}
   =\sum_\alpha \langle
\Omega,t_{\rm in}|\alpha,t_{\rm out}\rangle_{J_-}\langle\alpha,t_{\rm out}
         |\Omega,t_{\rm in}\rangle_{J_+},
\label{GeneratingFunctional}
\eeq
where the sum goes over a complete set of {\it out} states.
While in the standard {\it in-out} formalism one evolves the
{\it in} state in the presence of a source $J$ and compares it with the
{\it out} state, in the {\it in-in} framework
one evolves the {\it in} state in the presence of the two different sources
$J_+$ and $J_-$ and compares the results in the
remote future. In other words, expression~(\ref{GeneratingFunctional})
can be understood as the \emph{in}-vacuum going forward in time under
 influence of the source $J_+$ and then returning back in time under
the influence of the source $J_-$ (hence the term the
 \emph{closed-time-path formalism} broadly used in literature~\cite{Hu:1987}).
 From the path-integral representation of the {\it in-in} generating functional
(for details see \emph{e.g.}~\cite{Jordan:1986}),
 \beq
Z[J_-,J_+]=\int \mathcal D\varphi_+\mathcal D\varphi_-
  \exp\left({\imath S[\varphi_+]-\imath S[\varphi_-]
        +\imath\int d^DxJ_+ \varphi_+
        -\imath\int d^DxJ_-\varphi_-}\right)
 \label{in-in},
 \eeq
one obtains the expectation values upon differentiating
with respect to $J_+$ and $J_-$ and then setting $J_+=J_-$:
\bea
\frac{\delta}{-\imath\delta J_-(y_1)}\cdots
              \frac{\delta}{-\imath\delta J_-(y_m)}
 &\times& \frac{\delta}{\imath\delta J_+(x_1)}
    \cdots \frac{\delta}{\imath\delta J_+(x_n)} Z[J_-,J_+]_{J_-=J_+=0}=
\label{In-In ExpVal}\\ \nonumber
\qquad\qquad && \langle\Omega,t_{\rm in}|\{\overline T\varphi(y_1)
   \cdots\varphi(y_m)\}\times \{T\varphi(x_1)\cdots\varphi(x_n)\}
       |\Omega,t_{\rm in}\rangle,
\eea
with the boundary (initial) conditions for the {\it in}-vacuum state,
and the constraint for the fields on the future boundary
$\varphi_+(t_{out})=\varphi_-(t_{out})$.
In~(\ref{In-In ExpVal}) $T$ and $\overline T$
denote time and antitime ordering operators, \emph{i.e.},
 $T\varphi(x)\varphi(x^\prime)=\Theta(x^0-x^{\prime 0})\varphi(x)\varphi(x^\prime)+\Theta(x^{\prime 0}-x^{0})\varphi(x^\prime)\varphi(x)$ and $\overline T\varphi(x)\varphi(x^\prime)=\Theta(x^{\prime 0}-x^0)\varphi(x)\varphi(x^\prime)+\Theta(x^0-x^{\prime 0})\varphi(x^\prime)\varphi(x)$.

The generating functional is now an integral over two fields which we shall
conveniently combine into a column matrix:
\beq
\mathcal\varphi=
\left({\begin{array}{cc}
\varphi_+ \\
 \varphi_-\\
 \end{array} } \right),
\qquad
\mbox{with the source fields:}\qquad
\mathcal{J}=
\left({\begin{array}{cc}
 J_+ \\
 J_-\\
 \end{array} } \right).
\eeq
To calculate the generating functional $Z[J_-,J_+]$ is a rather formidable task, which principally
can be performed only in the perturbative setting.
Hence we decompose the action $S[\varphi]$ into its free
($S_F[\varphi]$) and interacting part ($S_I[\varphi]$):
\beq S[\varphi]=S_F[\varphi]+S_I[\varphi]\;,\qquad
  S_F[\varphi]=\int d^D x\frac 12 \varphi \mathcal D_0\varphi
\,,
\label{Action Decomposed}
\eeq
where for a nonminimally coupled scalar field
the kinetic operator $\mathcal D_0$ in~(\ref{Action Decomposed})
takes the form: ${\mathcal D}_0 = \Box-m^2-\xi R$,
where $\Box$ is the d'Alembertian operator, $m$ is the scalar field
mass, $R$ is the Ricci scalar and $D$ is the dimension of space-time.
The {\it in-in} generating functional~(\ref{in-in}) now becomes
\bea
Z[J_-,J_+]&=&{\rm e}^{\imath S_I^+[-\imath \delta/\delta J_+]
                        -\imath S_I^-[\imath \delta/\delta J_-]}
         \times \int \mathcal D\varphi_+\mathcal D\varphi_-\\ \nonumber
&\times&
\exp\bigg\{\int d^4 x\left[\frac 12(\varphi_+,\varphi_-)
\left({\begin{array}{cc}
\imath \mathcal{D}_0 & 0 \\
 0 & -\imath \mathcal{D}_0\\
 \end{array} } \right)
 \left( {\begin{array}{cc}
 \varphi_+\\
 \varphi_-\\
 \end{array}} \right)
 + (\varphi_+,\varphi_-)
\left({\begin{array}{cc}
\imath J_+  \\
 -\imath J_-\\
 \end{array} } \right)
 \right]\bigg\}
\,.
\eea
Next we shift the fields
\bea
 \left( {\begin{array}{cc}
 \varphi_+\\
 \varphi_-\\
 \end{array}} \right)
 \rightarrow
 \left( {\begin{array}{cc}
 \varphi_+\\
 \varphi_-\\
 \end{array}} \right)
 -
\left({\begin{array}{cc}
\imath \Delta_{++} & \imath \Delta_{+-} \\
\imath \Delta_{-+} & \imath \Delta_{--}\\
 \end{array} } \right)
 \left( {\begin{array}{cc}
 \imath J_+\\
 -\imath J_-\\
 \end{array}} \right)
\,,
\eea
which results in
\bea
Z[J_-,J_+]&=&e^{\imath S_I^+[-\imath \delta/\delta J_+]
                  -\imath S_I^-[\imath \delta/\delta J_-]}
\\ \nonumber
&\times&
\exp\bigg\{\int d^4x d^4x^\prime\frac 12(\imath J_+,-\imath J_-)(x)
 \left({\begin{array}{cc}\imath\Delta_{++} & \imath\Delta_{+-}\\
 \imath\Delta_{-+} & \imath\Delta_{--}\\
 \end{array}}\right)(x;x^\prime)
   \left({\begin{array}{cc}\imath J_+\\-\imath J_-\\
\end{array}}\right)(x^\prime)\bigg\},
 \label{In-In Propagators}
 \eea
 where the matrix of $\Delta$'s is a $2\times 2$ Keldysh matrix of (free)
propagators, defined as the
 inverse of $\mathcal D_0$'s:
\begin{equation}
\left( {\begin{array}{cc}
 \mathcal{D}_0 & 0 \\
 0 & -\mathcal{D}_0 \\
 \end{array} } \right)(x)
 \left( {\begin{array}{cc}
 \imath\Delta_{++} & \imath\Delta_{+-} \\
 \imath\Delta_{-+}&\imath\Delta_{--} \\
 \end{array} } \right)(x;x')=\imath\delta^D(x-x').
\end{equation}
Here $\imath\Delta_{++}$ is the free Feynman (time ordered),
$\imath\Delta_{--}$ is the free anti-Feynman (or Dyson) propagator and
$\imath\Delta_{+-}$ and $\imath\Delta_{-+}$ are the
(positive and negative frequency) free Wightman functions.
If we now insert Eq.~(\ref{In-In Propagators}) into 
Eq.~(\ref{In-In ExpVal}) we obtain relations
between expectation values and propagators:
\bea
\imath\Delta_{++}(x,x^\prime)
 &=&\left<\Omega|T\left(\varphi(x)\varphi(x^\prime)\right)|\Omega\right>
=\Theta(x^0-x^{\prime 0})\imath\Delta_{-+}(x;x^\prime)
  +\Theta(x^{\prime 0}-x^{0})i\Delta_{+-}(x;x^\prime),
\label{FeynmanPropagator}\\
\imath\Delta_{--}(x,x^\prime)
  &=&\left<\Omega|\overline{T}\left(\varphi(x)\varphi(x^\prime)\right)|\Omega\right>
=\Theta(x^0-x^{\prime 0})\imath\Delta_{+-}(x;x^\prime)
        +\Theta(x^{\prime 0}-x^{0})\imath\Delta_{-+}(x;x^\prime)
\,.
\label{AntiFeynmanPropagator}
\eea
It is worth noting that these relations hold both for free and dressed
propagators. When $\varphi$ is a massless minimally coupled scalar field,
\emph{i.e.} $\mathcal D_0=\partial^2$, the free propagator equations
reduce to
\bea
\partial^2 \imath\Delta_{++}(x;x^\prime)&=&\imath \delta^D(x-x^\prime)
\,,\qquad\quad
\partial^2 \imath\Delta_{+-}(x;x^\prime)=0\;,
\\ \nonumber
\partial^2 \imath\Delta_{--}(x;x^\prime)&=&-\imath \delta^D(x-x^\prime)
\,,\qquad\; \partial^2 \imath\Delta_{-+}(x;x^\prime)=0
\,.
\eea
The vacuum solution to all four propagators can be written compactly
as
\begin{equation}
\imath \Delta_{\pm\pm}(x;x')
  = \frac{\Gamma\left(\frac{D}{2}-1\right)}{4\pi^{D/2}}
 \left(\frac{1}{\Delta x^2_{\pm\pm}} \right)^{\frac{D}{2}-1}
\,,
\label{massless scalar propagators}
\end{equation}
where $\Delta x_{\pm\pm}$ are the four Schwinger-Keldysh length functions:
\bea
\Delta x_{++}^2&=&\|\vec{x}-\vec x^{\,\prime}\|^2
           -\left(| t-t'|-\imath\delta  \right)^2
,\qquad
    \Delta x_{+-}^2=\|\vec{x}-\vec x^{\,\prime}\|^2
      -\left(t-t'+\imath\delta  \right)^2\;,\\ \nonumber
\Delta x_{--}^2&=&\|\vec{x}-\vec x^{\,\prime}\|^2
   -\left(| t-t'|+\imath\delta  \right)^2,\qquad
  \Delta x_{-+}^2=\|\vec{x}-\vec x^{\,\prime}\|^2
       -\left(t-t'-\imath\delta  \right)^2
\,,
\label{distance functions}
\eea
where $\delta>0$ is an infinitesimal parameter, which defines
the time integration contours needed when propagators act on functions or
distributions.
Notice that $\Delta x_{++}^2$ and $\Delta x_{--}^2$ are Lorentz invariant,
such that the Feynman and anti-Feynman propagators
in~(\ref{massless scalar propagators}) are manifestly Lorentz invariant.

\section*{Appendix B: Classical Newtonian potential}
\label{Appendix B: Classical Newtonian potential}

In order to obtain the Einstein-Hilbert action to quadratic order
in $h_{\mu\nu}$ (see Sec. II) we first expand the Ricci scalar
in powers of $h_{\mu\nu}$. The general result is (see, for
example, Appendix~A of Ref.~\cite{Prokopec:2010be}),
\bea
 \sqrt{-g}R&=&\sqrt{-g}
      g^{\mu\nu}\bigg\{g^{\rho\sigma}[\partial_\rho\partial_\mu h_{\sigma\nu}
                        - \partial_\mu\partial_\nu h_{\rho\sigma}]
                   + g^{\rho\alpha}g^{\sigma\beta}
    \Big[-(\partial_\rho h_{\alpha\beta})(\partial_\mu h_{\sigma\nu})
      -\frac14(\partial_\sigma h_{\nu\alpha})(\partial_\mu h_{\beta\rho})
\nonumber\\
    && \hskip 1.5cm
      -\frac14(\partial_\sigma h_{\nu\alpha})(\partial_\rho h_{\mu\beta})
      -\frac14(\partial_\beta h_{\mu\nu})(\partial_\sigma h_{\rho\alpha})
      +(\partial_\sigma h_{\mu\nu})(\partial_\rho h_{\alpha\beta})
      +\frac34(\partial_\mu h_{\rho\sigma})(\partial_\nu h_{\alpha\beta})
    \Big]
             \bigg\},
\label{expanding Ricci} \eea
where
\begin{equation}
 \sqrt{-g} = 1 + \frac12 h
             + \Big(\frac18 h^2 - \frac14 h_{\mu\nu} h^{\mu\nu}\Big)
             + {\cal O}(h_{\mu\nu}^3)
\;,\qquad
  g^{\mu\nu} = \eta^{\mu\nu} - h^{\mu\nu} +  h^\mu_\rho h^{\rho\nu}
             + {\cal O}(h_{\mu\nu}^3)\,.
\label{metric:expansion in h}
\end{equation}
Notice that~(\ref{expanding Ricci}) contains all orders of
$h_{\mu\nu}$, and hence can be used to obtain the cubic, quartic
and higher order gravitational vertices. When Eqs.~(\ref{expanding
Ricci}--\ref{metric:expansion in h}) are combined,
 we get for the linear and quadratic contribution to the Einstein-Hilbert
action,
\bea \sqrt{-g}R &=&
            \Big(1+\frac{h}{2}\Big)
                       (\partial_\mu\partial_\nu h^{\mu\nu}-\partial^2 h)
           - h_{\rho\sigma}(2\partial^\rho\partial^\mu h^\sigma_{\;\mu}
                 -\partial^2 h^{\rho\sigma})
           + h^{\mu\nu}\partial_\mu\partial_\nu h
\nonumber\\
      &&   -\,\frac32(\partial_\rho h^{\rho\sigma})(\partial^\mu h_{\mu\sigma})
         -\frac14(\partial_\mu h)(\partial^\mu h)
        +(\partial^\mu h)(\partial^\nu h_{\mu\nu})
        +\frac34 (\partial^\mu h^{\nu\sigma})(\partial_\mu h_{\nu\sigma})
        + {\cal O}(h_{\mu\nu}^3)
\label{Ricci scalar:spatial:quadratic}
\\
 &=& -\frac14h\partial^2 h
        +\frac12 h\partial_\mu\partial_\nu h^{\mu\nu}
        -\frac12 h^{\mu\nu}\partial_\mu\partial_\sigma h_{\;\nu}^{\sigma}
        +\frac14 h^{\mu\nu}\partial^2 h_{\mu\nu}
        + {\cal O}(h_{\mu\nu}^3)
 + {\tt tot.\;der.\;terms}
\,, \nonumber \eea
where $\partial^2 = \eta^{\mu\nu}\partial_\mu\partial_\nu$,
$h^{\mu\nu} = \eta^{\mu\rho}\eta^{\nu\sigma}h_{\rho\sigma}$,
$\partial^\mu = \eta^{\mu\nu}\partial_\nu$,
$h=\eta^{\mu\nu}h_{\mu\nu}$. In the last line of~(\ref{Ricci
scalar:spatial:quadratic}) we dropped several total derivative
terms (which contribute as boundary terms, and hence do not
contribute to the equations of motion).
 The last line
in~(\ref{Ricci scalar:spatial:quadratic}) can be used to recover
the quadratic action~(\ref{Einstein-Hilbert:2}).

 Next, for pedagogical reasons here we present a derivation of
the classical potentials~(\ref{classical solution}), which solve
Eq.~(\ref{EOM cl}):
\beq
L_{\mu\nu\rho\sigma}h^{\rho\sigma(0)}(x)=\;\frac{\kappa^2}{2}\;\delta_\mu^0\delta_\nu^0
M\delta^{D-1}(\vec x),
 \label{App:EOM cl}
\eeq
where $\kappa^2 = 16\pi G_N$ and the Lichnerowicz operator is given by
\beq
L_{\mu\nu\rho\sigma}
 = \partial_{(\rho}\eta_{\sigma)(\mu}\partial_{\nu)}
  - \frac{1}{2}\eta_{\mu(\rho}\eta_{\sigma)\nu}\partial^2
  - \frac{1}{2}(\eta_{\rho\sigma}\partial_\mu\partial_\nu
                +\eta_{\mu\nu}\partial_{\rho}\partial_\sigma)
  +\frac12 \eta_{\mu\nu}\eta_{\rho\sigma}\partial^2
\,.
\label{App:Lichnerowicz}
\eeq
Let us first write the $(00)$, $(0i)$ and $(ij)$ components of
Eq.~(\ref{App:EOM cl}):
\bea
 \frac12(\partial_i\partial_j - \delta_{ij}\nabla^2)h_{ij}
            &=& \frac{\kappa^2}{2} M\delta^3(\vec x\,)
\label{App:EOM:00}\\
 -\frac12\nabla^2h_{0i}+\frac12\partial_i\partial_jh_{0j}
 +\frac12\partial_0(\partial_jh_{ij}-\partial_ih_{jj})
            &=& 0
\label{App:EOM:0i}\\
 \frac12 (\partial_i\partial_j-\delta_{ij}\nabla^2)h_{00}
            -\partial_0\partial_{(i}h_{j)0}
            +\delta_{ij}\partial_0\partial_lh_{0l}
            +\partial_l\partial_{(i}h_{j)l}
            -\frac12(-\partial_0^2+\nabla^2)h_{ij}
&&
\nonumber\\
            +\frac12\delta_{ij}(-\partial_0^2+\nabla^2)h_{ll}
            -\frac12(\delta_{ij}\partial_k\partial_lh_{kl}
                     +\partial_i\partial_jh_{ll})
            &=& 0
\,,
\label{App:EOM:ij}
\eea
where $\nabla^2 = \delta_{ij}\partial_i\partial_j \equiv \partial_i\partial_i$
and $h_{ij}$ are the spatial components of $h_{\mu\nu}$.

 Next we perform the standard scalar-vector-tensor decomposition of
 $h_{\mu\nu}$:
%
\beq
 h_{0i} = n_{i}^T+\partial_i \sigma
\;,\qquad
 h_{ij} = \frac{\delta_{ij}}{3}h
        +\Big(\partial_i\partial_j-\frac13\delta_{ij}\nabla^2\Big)\tilde h
        + \partial_i h_j^T + \partial_j h_i^T + h_{ij}^{TT}
\,,
\label{App:SVT decomposition}
\eeq
such that $\partial_i n_i^T = 0 = \partial_i h_i^T$, $h_{ii}^{TT}=0$,
 $\partial_i h_{ij}^{TT}=0$ and ${\rm Tr}[h_{ij}]=h$.
Consider now an infinitesimal coordinate
transformation, $x^\mu\rightarrow x^\mu + \xi^\mu(x)$.
Then, as a consequence of
the tensorial transformation law,
the metric perturbation on Minkowski background transforms as,
\beq
 h_{\mu\nu}(x) \rightarrow h_{\mu\nu}(x) - \partial_\mu \xi_\nu(x)
                                   - \partial_\mu \xi_\nu(x)
                                   +{\cal O}(\xi^2,h^2,\xi h)
\,,
\label{App:transformation}
\eeq
(with $\xi_\mu = \eta_{\mu\nu}\xi^\nu$) where it 
is also known as the gauge transformation of general relativity.
In the light of the metric decomposition~(\ref{App:SVT decomposition}),
it is also convenient
to decompose $\xi^\mu$ into two scalars and a transverse vector as 
$\xi_\mu = (\xi_0,\xi^T_i,\partial_i\xi)$, where $\partial_i\xi^T_i=0$.
With this~(\ref{App:transformation}) can be rewritten in components as
\bea
  h_{00}&\rightarrow& h_{00} -2\partial_0\xi_0
\;,\qquad
  n_i^T\rightarrow n_i^T -\partial_0\xi_i^T
\;,\qquad
  \sigma\rightarrow \sigma -\partial_0\xi-\xi_0\;,
\nonumber\\
  h_{ij}^{TT}&\rightarrow& h_{ij}^{TT}
\;,\qquad
  h_i^T\rightarrow h_i^T -\xi_i^T
\;,\qquad
  \tilde h\rightarrow \tilde h  -2\xi
\;,\qquad
   h\rightarrow h - 2\nabla^2\xi
\,.
\label{App:transformation:2}
\eea
From these relations it then immediately follows that the following
two scalars, one vector and the tensor are gauge invariant:
\bea
  h_{00}-2\partial_0\sigma+\partial_0^2\tilde h
\;,\qquad
   h- \nabla^2\tilde h
\;,\qquad
  n_i^T - \partial_0h_i^T
\;,\qquad
  h_{ij}^{TT}
\,.
\label{App:transformation:gi}
\eea
 This means that out of the ten components of $h_{\mu\nu}$, six are gauge
invariant (and correspond to observables), while the remaining four components
are gauge dependent, and have no independent physical meaning.
It is now instructive to
rewrite Eqs.~(\ref{App:EOM:00}--\ref{App:EOM:ij}) in terms of the scalar,
vector and tensor components of $h_{\mu\nu}$ defined
in~(\ref{App:SVT decomposition}). The resulting equations are
\bea
 \nabla^2(h-\nabla^2\tilde h)
            &=& -\frac{3\kappa^2}{2} M\delta^3(\vec x\,)\;,
\label{App:EOM:00:2}\\
\nabla^2 (n_i^T - \partial_0h_i^T) &=& 0 \;,\qquad
\partial_0\partial_i(h- \nabla^2\tilde h) = 0\;,
\label{App:EOM:0i:2}\\
(\partial_i\partial_j-\delta_{ij}\nabla^2)
         \Big[\frac12(h_{00}-2\partial_0\sigma+\partial_0^2\tilde h)
              - \frac16(h-\nabla^2\tilde h)\Big] &=& 0
\;,\quad\;
            \partial_0^2(h-\nabla^2\tilde h) = 0
\;,\quad\;
             \partial_0\partial_i(n_j^T-\partial_0 h_j^T) = 0\;,
\label{App:EOM:ij:2}\;\;\;\quad
\\
            (-\partial_0^2+\nabla^2)h_{ij}^{TT} &=& 0
\,,
\label{App:EOM:ijTT:2}
\eea
 These equations possess a number of remarkable properties.
First, Eq.~(\ref{App:EOM:ijTT:2}) is the only dynamical equation
and it is the wave equation for gravitational waves, obeying the
standard Lorentz covariant wave operator. Second, all equations
are gauge invariant (see Eq.~(\ref{App:transformation:gi})), and
hence can be written in the gauge invariant form as
\beq \nabla^2 \Psi = 4\pi G_NM\delta^3(\vec x\,) \;,\qquad
\partial_0^2 \Psi = \partial_0\partial_i \Psi = 0
\,,\qquad (\partial_i\partial_j-\delta_{ij}\nabla^2)(-\Phi +
\Psi)=0 \;,\qquad
\partial_0\partial_i \tilde n_i^T = \nabla^2 \tilde n_i^T = 0
\,,
\label{App:EOM:gi}
\eeq
where we introduced the usual Bardeen potentials,
\beq \Psi = -\frac16(h-\nabla^2\tilde h) \;,\qquad
 \Phi =  -\frac12(h_{00}-2\partial_0\sigma+\partial_0^2\tilde
 h)\;,
\label{App:Bardeen potentials}
\eeq
and a gauge invariant shift vector,
\beq
 \tilde n_i^T = n_i^T - \partial_0 h_i^T
\,.
\label{vector:gi}
\eeq
Eqs.~(\ref{App:EOM:gi}) can be easily solved assuming that
$\Phi-\Psi$ and $\tilde n_i^T$ vanish at spatial infinity:
\beq
 \Phi = \Psi = -\frac{G_NM}{r}
\;,\qquad \tilde n_i^T = 0 \,, \label{App:EOM:solutions:gi} \eeq
where $r = \|\vec x\,\|$ and we made use of
 the Green function $G(\vec x\,)=-1/(4\pi r)$ for
$\nabla^2G(\vec x\,) = \delta^3(\vec x\,)$. Notice that there are
3 equations for $\Psi$ in~(\ref{App:EOM:gi}). These equations are
not redundant. In fact, they assure that the linearized version of
the Birkhoff theorem holds: the potentials of linearized gravity
generated by a static point mass must be time independent.
Notice finally that, the linearized
Eqs.~(\ref{App:EOM:gi}--\ref{App:EOM:ijTT:2}) do not tell us
anything about the gauge variant quantities (among which are two
scalars and one vector), which is also what one expects.

 Most of the literature does not make use of the (gauge invariant)
Bardeen potentials to specify the gravitational response to a
static point mass, but instead one fixes a gauge. The main purpose
of this Appendix, in which we show how to construct classical
gauge invariant potentials, is to facilitate comparison with the 
literature, which uses a variety of gauges. For example, in the
longitudinal (Newton) gauge, the scalar metric sector is fixed to
be
\begin{equation}
 ds^2 = h_{00}dt^2 + \sum_{i=1}^3 h_{ii}(d x^i)^2
\,.
\label{longitudinal gauge}
\end{equation}
In this gauge we have simply
\beq
 \Phi=-\frac12 h_{00} \;,\qquad
 \Psi=-\frac16{\rm Tr}[h_{ij}]=-\frac12 h_{ii} \;\;\;({\rm no \; summation})
\,.
\label{PhiPsi:long gauge}
\eeq
Together with~(\ref{App:EOM:solutions:gi}), these equations
imply~(\ref{classical solution}), which is the classical Newtonian potential
that we use in this work.

On the other hand, Park and Woodard~\cite{Woodard:2010}
make use of the non-diagonal gauge,
\beq
 h_{00} = \frac{2G_NM}{r}
\;,\qquad
 h_{ij} = \frac{2G_NM}{r}\hat r_i\hat r_j
\,.
\label{ParkWoodard gauge}
\eeq
This gauge can be obtained from the two scalars of $h_{ij}$
in~(\ref{App:SVT decomposition}) by observing
that for a radial scalar function $\tilde h=\tilde h(r)$,
$\partial_i  = \hat r_id/dr$ ($\hat r_i=x^i/r$) and
$\partial_i\partial_j = (\delta_{ij}/r)(d/dr)
 +\hat r_i\hat r_j[(d^2/dr^2)-(1/r)(d/dr)]$, from which
it follows,
\beq
 (h_{ij})_{\rm scalar} = \frac{\delta_{ij}}{3}h
                       + \Big(\partial_i\partial_j-\frac13 \delta_{ij}\nabla^2
                             \Big)\tilde h
      =\frac{\delta_{ij}}{3}
          \Big[h-\Big(\frac{d^2}{dr^2}-\frac1r\frac{d}{dr}\Big)\tilde h\Big]
      +\hat r_i\hat r_j \Big(\frac{d^2}{dr^2}-\frac1r\frac{d}{dr}\Big)\tilde h
\,.
\label{ParkWoodard gauge:2}
\eeq
Comparing this with~(\ref{ParkWoodard gauge}) reveals that
in that gauge
\beq
  h =\Big(\frac{d^2}{dr^2}-\frac1r\frac{d}{dr}\Big)\tilde h = \frac{2G_NM}{r}
\,.
\label{ParkWoodard gauge:3}
\eeq
 To check these relations, note first that, up to a constant,
$\tilde h = -2G_NMr$. Equation~(\ref{App:Bardeen potentials}) then implies
\beq
\Psi=-\frac16\Big[h-\Big(\frac{d^2}{dr^2}+\frac2r\frac{d}{dr}\Big)\tilde h\Big]
      = -\frac{G_NM}{r}
\,,
\label{ParkWoodard gauge:4}
\eeq
which agrees with~(\ref{App:EOM:solutions:gi}), implying
that Eqs.~(\ref{ParkWoodard gauge}) represent a correct classical solution.

\section*{Appendix C: Expanding the bubble diagram}
\label{Appendix C: Expanding the bubble diagram}

 In this Appendix we expand the scalar field bubble diagram
in Eq.~(\ref{2pi:1loop}) shown in Fig.~\ref{figure 3} in powers of small
(quantum) metric perturbations $\delta g^{\mu\nu}(x)
= -h^{\mu\nu}(x) + h^{\mu\mu^\prime}(x)h_{\mu^\prime}^{\;\nu}(x)
 +{\cal O}((h^{\mu\nu})^3)$.

Let us begin by noting that the scalar propagator in~(\ref{2pi:1loop}) obeys
the equation
\beq
 (-g^{(b)}(x))^{1/2}\Box^{(b)}_x
    \Delta^{ab}(x;x';[g^{(b)}_{\alpha^\prime\beta^\prime}])
  = \partial_\mu {g^{(b)}}^{\mu\nu}(x)[-g^{(b)}(x)]^{1/2}\partial_\nu
             \Delta^{ab}(x;x';[g^{(b)}_{\alpha^\prime\beta^\prime}])
            = (\sigma^3)^{ab}\delta^D(x-x^\prime\,)
\,,
\label{scalar propagator equation}
\eeq
such that $\sqrt{-g^{(b)}(x)}\Box^{(b)}_x\delta^D(x-y)(\sigma^3)^{ab}$
is the operator inverse of
$\Delta^{ab}(y;x';[g^{(b)}_{\alpha^\prime\beta^\prime}])$.
$\sigma^3={\rm diag}(1,-1)$ in~(\ref{scalar propagator equation})
is the Pauli matrix.

 Writing $g^{(b)}_{\mu\nu}=g_{\mu\nu}+\delta g_{\mu\nu}$, we can expand
the bubble diagram in~(\ref{2pi:1loop}) in powers of $\delta
g_{\mu\nu}$. To proceed, let us vary the first bubble diagram
in~(\ref{2pi:1loop}) with respect to $g^{\mu\nu}(x)$ (for
notational simplicity we drop the Keldysh indices $a,b,..=\pm$ and
the background index $(b)$):
\bea
\frac{\delta}{\delta g^{\mu\nu}(x)}\bigg[-\frac{\imath}{2}{\rm Tr}
                      \ln[\Delta_F(y;y;[g^{(b)}_{\alpha^\prime\beta^\prime}])]
                                  \bigg]
 &=& \frac{\delta}{\delta g^{\mu\nu}(x)}
       \bigg[\frac{\imath}{2}\int d^Dy\Big\{
        \ln[\sqrt{-g(y)}\Box_y\delta^D(y-y^\prime\,)|_{y^\prime\rightarrow y}]
                                      \Big\}
       \bigg]
\label{2pi:1loop:2}
\\
&&\hskip -5.2cm
=\, \frac{\imath}{2}\int d^Dy\int d^Dz
        \partial_\alpha^y
          \Big[\Big(-\frac12\sqrt{-g(y)}g^{\alpha\beta}(y)g_{\mu\nu}(y)
              +\sqrt{-g(y)}\delta^\alpha_{(\mu}\delta^\beta_{\nu)}\Big)
                  \delta^D(y-x)\partial^y_\beta\delta^D(y-z)\Big]
                            \Delta_F(z;y;[g_{\alpha^\prime\beta^\prime}])
\,,
\nonumber
\eea
where $\Delta_F(z;y;[g_{\alpha^\prime\beta^\prime}])
   =\Delta^{++}(z;y;[g_{\alpha^\prime\beta^\prime}])$
 denotes the Feynman propagator.
Now, because of the $\delta^D(y-x)$ function, all of the $y$'s
in the square brackets can be replaced by $x$'s, and a part can be pulled
out of the integral to obtain
\bea
\Big(-\frac12\sqrt{-g(x)}g^{\alpha\beta}(x)g^{\mu\nu}(x)
              +\sqrt{-g(x)}\delta^\alpha_{(\mu}\delta^\beta_{\nu)}\Big)
      \frac{\imath}{2}\int d^Dy\int d^Dz
                  \delta^D(y-x)\big[\partial^x_\beta\delta^D(x-z)\big]
                            \Delta_F(z;y;[g_{\alpha^\prime\beta^\prime}])
                      \big[\partial_\alpha^y\delta^D(y-x)\big]
\,.
\nonumber
\eea
Next, the derivatives
$\partial^x_\beta\delta^D(x-z)=-\partial^z_\beta\delta^D(x-z)$ and
$\partial_\alpha^y$ can be moved by partial integration to act on
the propagator $\Delta_F(z;y[g_{\alpha\beta}])$, and finally the
two integrals can be performed on the expense of the
$\delta$-functions. The result is
\bea
\frac{\delta}{\delta g^{\mu\nu}(x)}\bigg[\!-\!\frac{\imath}{2}{\rm Tr}
                      \ln[\Delta_F(y;y;[g_{\alpha^\prime\beta^\prime}])] \bigg]
 \!&=&\!
\frac{1}{2}
\Big(\frac12\sqrt{-g(x)}g^{\alpha\beta}(x)g_{\mu\nu}(x)
              \!-\!\sqrt{-g(x)}\delta^\alpha_{(\mu}\delta^\beta_{\nu)}\Big)
\Big[\partial_\alpha^x\partial^y_\beta
         \imath\Delta_F(x;y;[g_{\alpha^\prime\beta^\prime}])
 \Big]_{y\rightarrow x}
.\quad\;\,\,
\label{2pi:1loop:3}
\eea
 This is the general expression for the tadpole diagram,
{\it i.e.} the linear coupling of the stress energy tensor to metric
perturbations.

 In order to get the vacuum polarization
we need to vary~(\ref{2pi:1loop:3}) one more time
with respect to $g^{\rho\sigma}(x^\prime)$. To facilitate the variation
procedure, recall the operator inverse identity:
\bea
 \int d^D z \partial^{z^\prime}_\gamma \sqrt{-g(z^\prime)}g^{\gamma\delta}(z^\prime)
      \partial_\delta^{z^\prime}
    \delta^D(z^\prime-z)\Delta_F(z;y;[g_{\alpha^\prime\beta^\prime}])
              =  \delta^D(z^\prime-y)
\,.
\nonumber
\eea
Varying this  with respect to $g^{\rho\sigma}(x^\prime)$ yields
\bea
&&\hskip -0.5cm \int d^D z\int d^D z^\prime
           \Delta_F(x;z^\prime;[g_{\alpha^\prime\beta^\prime}])
      \partial^{z^\prime}_\gamma \Big[\Big(
        \!-\!\frac12\sqrt{-g(z^\prime)}g^{\gamma\delta}(z^\prime)
                  g_{\rho\sigma}(z^\prime)
               \delta^D(z^\prime-x^\prime)
        +\sqrt{-g(z^\prime)}\delta^\gamma_{(\rho}\delta^\delta_{\sigma)}
               \delta^D(z^\prime-x^\prime)\Big)
    \partial^{z^\prime}_\delta\delta^D(z^\prime\!-\!z)\Big]
\nonumber\\
&&\hskip 4cm \times\,\Delta_F(z;y;[g_{\alpha^\prime\beta^\prime}])
 + \int d^Dz\delta^D(x\!-\!z)
       \frac{\delta}{\delta g^{\rho\sigma}(x^\prime)}
   \Delta_F(z;y;[g_{\alpha^\prime\beta^\prime}])
 =  0
\,.
\nonumber
\eea
Setting $y\rightarrow x$ and performing similar steps as above,
this expression can be recast as
\bea
\bigg[\frac{\delta}{\delta g^{\rho\sigma}(x^\prime)}
    \Delta_F(x;y;[g_{\alpha^\prime\beta^\prime}])
  \bigg]_{y\rightarrow x}
 &=& -\Big(\frac12\sqrt{-g(x^\prime)}g^{\gamma\delta}(x^\prime)g_{\rho\sigma}(x^\prime)
        -\sqrt{-g(x^\prime)}\delta^\gamma_{(\rho}\delta^\delta_{\sigma)}
   \Big)
\nonumber\\
 &&\hskip 0.4cm\times\, \Big[\partial_\gamma^{x^\prime}
     \Delta_F(x;x^\prime;[g_{\alpha^\prime\beta^\prime}])\Big]
\Big[\partial_\delta^{x^\prime}
     \Delta_F(x^\prime;x;[g_{\alpha^\prime\beta^\prime}])\Big]
\,.
\label{2pi:1loop:4}
\eea
Inserting this into~(\ref{2pi:1loop:3}) we get for the second
variation of the bubble diagram:
\bea
\frac{\delta}{\delta g^{\rho\sigma}(x^\prime)}
  \frac{\delta}{\delta g^{\mu\nu}(x)}\bigg[-\frac{\imath}{2}{\rm Tr}
                      \ln[\Delta_F(y;y;[g_{\alpha^\prime\beta^\prime}])] \bigg]
 &=&
\, \frac{\imath}{2}
\sqrt{-g(x)}\Big(\frac12g^{\alpha\beta}(x)g_{\mu\nu}(x)
              -\delta^\alpha_{(\mu}\delta^\beta_{\nu)}\Big)
\Big[\partial_\alpha^x\partial_\gamma^{x^\prime}
 \imath \Delta_F(x;x^\prime;[g_{\alpha^\prime\beta^\prime}])\Big]
\nonumber\\
&&\hskip -3cm\times\,
\Big[\partial_\beta^x\partial_\delta^{x^\prime}
    \imath\Delta_F(x^\prime;x;[g_{\alpha^\prime\beta^\prime}])\Big]
 \sqrt{-g(x^\prime)}\Big(\frac12g^{\gamma\delta}(x^\prime)g_{\rho\sigma}(x^\prime)
        -\delta^\gamma_{(\rho}\delta^\delta_{\sigma)}
   \Big)
\nonumber\\
&&\hskip -5cm
 -\,\sqrt{-g(x)}\Big[\frac18g_{\rho\sigma}(x)g^{\alpha\beta}(x)g_{\mu\nu}(x)
              -\frac14\delta^\alpha_{(\rho}\delta^\beta_{\sigma)}g_{\mu\nu}(x)
              -\frac14\delta^\alpha_{(\mu}\delta^\beta_{\nu)}g_{\rho\sigma}(x)
              +\frac14g^{\alpha\beta}(x)g_{\mu(\rho}(x)g_{\sigma)\nu}(x)
              \Big]
\nonumber\\
&&\hskip -3.5cm\times\,\delta^D(x-x^\prime)
  \Big[\partial_\alpha^x\partial_\beta^{x^\prime}
           \imath\Delta_F(x;x^\prime;[g_{\alpha^\prime\beta^\prime}])\Big]
\,.
\label{2pi:1loop:5}
\eea
The last two lines come from varying the metric tensors in~(\ref{2pi:1loop:3}).
 Putting everything together, we get the following expansion of the
scalar bubble diagram:
\bea
-\frac{\imath}{2}{\rm Tr}\ln[\Delta_F(y;y;[g^{(b)}_{\alpha\beta}])]
&=&-\frac{\imath}{2}{\rm Tr}\ln[\Delta_F(y;y;[g_{\alpha^\prime\beta^\prime}])]
  +\frac12\int d^Dx\sqrt{-g(x)}\delta g^{\mu\nu}(x)
    T_{\mu\nu}(x;[g_{\alpha^\prime\beta^\prime}])
\label{bubble digram:expansion}
\\
 &-& \frac 12 \int d^Dx \sqrt{-g(x)}\int d^Dx^\prime \sqrt{-g(x^\prime)}
     \delta g^{\mu\nu}(x)
[\phantom{!}_{\mu\nu}
  \tilde\Pi_{\rho\sigma}](x;x^\prime;[g_{\alpha^\prime\beta^\prime}])
     \delta g^{\rho\sigma}(x^\prime)
 + {\cal O}\big((\delta g^{\alpha\beta})^3\big)
\,,
\nonumber
\eea
where
\bea
T_{\mu\nu}(x;[g_{\alpha^\prime\beta^\prime}])
   &=&
 \Big(\frac12g^{\alpha\beta}(x)g_{\mu\nu}(x)
              -\delta^\alpha_{(\mu}\delta^\beta_{\nu)}\Big)
\Big[\partial_\alpha^x\partial^{x^\prime}_\beta
  \imath\Delta_F(x;x^\prime;[g_{\alpha^\prime\beta^\prime}])
 \Big]_{x^\prime\rightarrow x}
\label{T:general}\\
\left[\phantom{!}_{\mu\nu}\tilde\Pi_{\rho\sigma}\right]
         (x;x^\prime;[g_{\alpha^\prime\beta^\prime}])
   &=& -\frac{\imath}{2}
\Big(\frac12g^{\alpha\beta}(x)g_{\mu\nu}(x)
              -\delta^\alpha_{(\mu}\delta^\beta_{\nu)}\Big)
\Big(\frac12g^{\gamma\delta}(x^\prime)g_{\rho\sigma}(x^\prime)
        -\delta^\gamma_{(\rho}\delta^\delta_{\sigma)}
   \Big)
\nonumber\\
&&\times\,
\Big[\partial_\alpha^x\partial_\gamma^{x^\prime}
 \imath \Delta_F(x;x^\prime;[g_{\alpha^\prime\beta^\prime}])\Big]
\Big[\partial_\beta^x\partial_\delta^{x^\prime}
    \imath\Delta_F(x^\prime;x;[g_{\alpha^\prime\beta^\prime}])\Big]
 \nonumber\\
&&+\,\Big[\frac18g_{\rho\sigma}(x)g^{\alpha\beta}(x)g_{\mu\nu}(x)
              -\frac14\delta^\alpha_{(\rho}\delta^\beta_{\sigma)}g_{\mu\nu}(x)
              -\frac14\delta^\alpha_{(\mu}\delta^\beta_{\nu)}g_{\rho\sigma}(x)
              +\frac14g^{\alpha\beta}(x)g_{\mu(\rho}(x)g_{\sigma)\nu}(x)
              \Big]
\nonumber\\
&&\times\,
\frac{\delta^D(x-x^\prime)}{\sqrt{-g(x)}}
  \Big[\partial_\alpha^x\partial_\beta^{x^\prime}
           \imath\Delta_F(x;x^\prime;[g_{\alpha^\prime\beta^\prime}])\Big]
\,.
\label{Sigma:general}
\eea
The interpretation of this result is simple.
The first term in~(\ref{bubble digram:expansion}) is the scalar 1PI bubble
in a classical background $g_{\alpha\beta}$, the second term
is the 1PI tadpole, with the metric perturbation
$\delta g^{\mu\nu}(x)=-h^{\mu\nu}(x)
= -g^{\mu\alpha}(x)g^{\nu\beta}(x)h_{\alpha\beta}(x)$ and
the matter stress energy tensor
$T_{\mu\nu}$ given in~(\ref{T:general}). The last
 term in~(\ref{bubble digram:expansion}) is the contribution
that is quadratic in the metric perturbation
$\delta g^{\mu\nu}\simeq -h^{\mu\nu}$,
with $\phantom{!}_{\mu\nu}
      \tilde\Pi_{\rho\sigma}(x;x^\prime;[g_{\alpha^\prime\beta^\prime}])$
being (a part of) the graviton vacuum polarization (calculated on
a $g_{\alpha\beta}(x)$ background), whose explicit form is given
in~(\ref{Sigma:general}). The first two lines in
(\ref{Sigma:general}) represent the nonlocal part of the
graviton vacuum polarization shown by the middle diagram in
Fig.~\ref{figure 2}, and they perfectly agree with the graviton
vacuum polarization shown in the first line of
Eq.~(\ref{Self-En-Big}) in Sec.~\ref{Graviton vacuum
polarization} ($\xi\rightarrow 0$ and $g_{\alpha\beta}\rightarrow
\eta_{\alpha\beta}$). The last two lines in~(\ref{Sigma:general})
represent a part of the local contribution to the vacuum
polarization. The other part is obtained by observing that the
quadratic term in the expansion $\delta
g^{\mu\nu}(x)=-h^{\mu\nu}(x)+h^\mu_\rho(x)h^{\rho\nu}(x) +{\cal
O}((h^{\mu\nu}(x))^3)$ in the tadpole in (\ref{bubble
digram:expansion}) yields the local contribution to the vacuum
polarization of the form
\bea
  \delta[\,_{\mu\nu}\Pi_{\rho\sigma}]
                   (x;x^\prime;[g_{\alpha^\prime\beta^\prime}])
  = \Big[-\frac12g^{\alpha\beta}(x)g_{\nu(\rho}g_{\sigma)\mu}(x)
         +\delta^\alpha_{(\mu}g_{\nu)(\rho}\delta^\beta_{\sigma)}\Big]
         \delta^D(x-x^\prime\,)
          \Big[\partial^x_\alpha\partial^{x^\prime}_\beta
               \imath \Delta_F(x;x^\prime;[g_{\alpha^\prime\beta^\prime}])\Big]
\,.
\label{tadpole to Sigma}
\eea
such that
\beq
 [\,_{\mu\nu}\Pi_{\rho\sigma}](x;x^\prime;[g_{\alpha^\prime\beta^\prime}])
 =[\,_{\mu\nu}\tilde\Pi_{\rho\sigma}](x;x^\prime;[g_{\alpha^\prime\beta^\prime}])
  +\delta[\,_{\mu\nu}\Pi_{\rho\sigma}](x;x^\prime;[g_{\alpha^\prime\beta^\prime}])
\label{full sigma from bubble}
\eeq
represents the complete graviton vacuum polarization, which
correctly includes also the local contribution from the quartic
interaction~(\ref{Squartic}) (in the limit when
$g_{\mu\nu}\rightarrow\eta_{\mu\nu}$ and $\xi\rightarrow 0$).
 A technique that is equivalent to expanding the 2PI bubble
diagram around the Minkowski background was used to calculate the leading quantum
correction to the Newtonian potential of a static point particle 
in Refs.~\cite{Dalvit:1994gf,Satz:2004hf}.
In the limit when time transients are
gone, our results presented in
Sec.~\ref{The Quantum Corrected Newtonian potential}
indeed agree with those of Refs.~\cite{Dalvit:1994gf,Satz:2004hf}, just as one
would expect from the analysis presented in this Appendix.

 And finally one more remark. From
Eqs.~(\ref{bubble digram:expansion}--\ref{T:general}) and the
preceding analysis it follows that the nonlocal part of the
graviton vacuum polarization is obtained by
\beq
\left\{ \left[\phantom{!}_{\mu\nu}\Pi_{\rho\sigma}\right](x;x^\prime)
 \right\}_{\rm non-local}
 = \frac{1}{\sqrt{g(x^\prime)}}
  \frac{\tilde\delta}{\tilde\delta g_{\rho\sigma}(x^\prime)}
     \left[T_{\mu\nu}(x;[g_{\alpha^\prime\beta^\prime}])\right]
\,,
\eeq
where the tilde on the functional derivative ($\tilde\delta$)
signifies that it acts on the propagator(s) in $T_{\mu\nu}$ only.
A special case of this relation can be found in Eqs.~(12-15)
and~(47) of Ref.~\cite{Woodard:2005}, where the non-local part of
the graviton vacuum polarization tensor around the Minkowski vacuum is
expressed in terms of the non-local, Wick contracted,
stress-energy--stress-energy correlator.

\section*{Appendix D: Non-local contributions}

In this Appendix we carry out thorough calculation of the graviton
vacuum polarization. By Wick-contracting all non-coincident
scalars and making use of the propagator
definitions~(\ref{FeynmanPropagator}--\ref{AntiFeynmanPropagator}),
the graviton vacuum polarization inferred from~(\ref{GammaCubic})
becomes
\bea
\imath\left[{}^{\pm}_{\mu\nu}\Pi_{\rho\sigma}^{\pm}\right](x;x^\prime)
&=&\frac{\pm\kappa}{2}\frac{\pm \kappa}{2} \bigg\{
2\left(\frac{1}{2}\eta^{\alpha\beta}\eta_{\mu\nu}
  \!-\!\delta_{(\mu}^{(\alpha}\delta_{\nu)}^{\beta)}\right)
\left[\partial_\alpha\partial^\prime_{(\alpha^\prime}
   \imath\Delta_{\pm\pm}(x;x^\prime)
  \partial^\prime_{\beta^\prime)}\partial_\beta
\imath\Delta_{\pm\pm}(x;x^\prime) \right]
\left(\frac{1}{2}\eta^{\alpha^\prime\beta^\prime}\eta_{\rho\sigma}
  \!-\!\delta_{(\rho}^{(\alpha^\prime}\delta_{\sigma)}^{\beta^\prime)}\right)
\nonumber
\\
\nonumber
&+&4\xi\left(\frac{1}{2}\eta^{\alpha\beta}\eta_{\mu\nu}
  -\delta_{(\mu}^{(\alpha}\delta_{\nu)}^{\beta)}\right)
\Big[
\partial_\alpha\partial^\prime_{(\rho}\imath\Delta_{\pm\pm}(x;x^\prime)
 \partial^\prime_{\sigma)}\partial_\beta\imath\Delta_{\pm\pm}(x;x^\prime)
\\
\nonumber
&&\hskip 1.5cm
-\,\eta^{\gamma^\prime\delta^\prime}\eta_{\rho\sigma}
\partial_\alpha\partial^\prime_{(\gamma^\prime}
 \imath\Delta_{\pm\pm}(x;x^\prime)
 \partial^\prime_{\delta^\prime)}\partial_\beta
\imath\Delta_{\pm\pm}(x;x^\prime)
+\partial_{(\alpha}\imath\Delta_{\pm\pm}(x;x^\prime)
  \partial_{\beta)}D^\prime_{\rho\sigma}
  \imath\Delta_{\pm\pm}(x;x^\prime)
 \Big]
\\ \nonumber
&+&4\xi \Big[
\partial^\prime_{\alpha^\prime}\partial_{(\mu}
   \imath\Delta_{\pm\pm}(x;x^\prime)
  \partial_{\nu)}\partial^\prime_{\beta^\prime}
   \imath\Delta_{\pm\pm}(x;x^\prime) -\eta^{\gamma\delta}\eta_{\mu\nu}
\partial^\prime_{\alpha^\prime}\partial_{(\gamma}
  \imath\Delta_{\pm\pm}(x;x^\prime)
 \partial_{\delta)}\partial^\prime_{\beta^\prime}
\imath\Delta_{\pm\pm}(x;x^\prime)
\\
\nonumber &&\qquad\qquad
+\,\partial^\prime_{(\alpha^\prime}\imath\Delta_{\pm\pm}(x;x^\prime)
  \partial^\prime_{\beta^\prime)}
  D_{\mu\nu}\imath\Delta_{\pm\pm}(x;x^\prime)\Big]
\,\left(\frac{1}{2}\eta^{\alpha^\prime\beta^\prime}\eta_{\rho\sigma}
  -\delta_{(\rho}^{(\alpha^\prime}\delta_{\sigma)}^{\beta^\prime)}\right)
\\
\nonumber
&+&\,8\xi^2\Big[
\partial_\mu\partial^\prime_{(\sigma}
  \imath\Delta_{\pm\pm}(x;x^\prime)
  \partial^\prime_{\rho)}\partial_\nu \imath\Delta_{\pm\pm}(x;x^\prime)
-\eta^{\gamma^\prime\delta^\prime}\eta_{\rho\sigma}
\partial_\mu\partial^\prime_{(\gamma^\prime}
   \imath\Delta_{\pm\pm}(x;x^\prime)
  \partial^\prime_{\delta^\prime)}\partial_\nu\imath\Delta_{\pm\pm}(x;x^\prime)
\\
\nonumber
&&\hskip -0.4cm
 -\,\eta^{\gamma\delta}\eta_{\mu\nu}
\partial_\gamma\partial^\prime_{(\rho}
  \imath\Delta_{\pm\pm}(x;x^\prime)
 \partial^\prime_{\sigma)}\partial_\delta\imath\Delta_{\pm\pm}(x;x^\prime)
\!+\!\eta^{\gamma\delta}\eta_{\mu\nu}
\eta^{\gamma^\prime\delta^\prime}\eta_{\rho\sigma}
\partial_\gamma\partial^\prime_{(\delta^\prime}
  \imath\Delta_{\pm\pm}(x;x^\prime)
\partial^\prime_{\gamma^\prime)}\partial_\delta\imath\Delta_{\pm\pm}(x;x^\prime)
\\
\nonumber
&&\quad +\,\partial_{(\mu}\imath\Delta_{\pm\pm}(x;x^\prime)
  \partial_{\nu)}D^\prime_{\rho\sigma}\imath\Delta_{\pm\pm}(x;x^\prime)
-\eta^{\gamma\delta}\eta_{\mu\nu}\partial_{(\gamma}
  \imath\Delta_{\pm\pm}(x;x^\prime)\partial_{\delta)}D^\prime_{\rho\sigma}
    \imath\Delta_{\pm\pm}(x;x^\prime)
\\
\nonumber
&&\qquad
+\,\partial^\prime_{(\rho}\imath\Delta_{\pm\pm}(x;x^\prime)
   \partial^\prime_{\sigma)}D_{\mu\nu}\imath\Delta_{\pm\pm}(x;x^\prime)
-\eta^{\gamma^\prime\delta^\prime}\eta_{\rho\sigma}\partial^\prime_{(\gamma^\prime}
  \imath\Delta_{\pm\pm}(x;x^\prime)\partial^\prime_{\delta^\prime)}D_{\mu\nu}
  \imath\Delta_{\pm\pm}(x;x^\prime)
\\
&&\qquad\;\;
+\,\frac{1}{2}\imath\Delta_{\pm\pm}(x;x^\prime)D_{\mu\nu}D^\prime_{\rho\sigma}
   \imath\Delta_{\pm\pm}(x;x^\prime)
+\frac{1}{2}D_{\mu\nu}\imath\Delta_{\pm\pm}(x;x^\prime)D^\prime_{\rho\sigma}
  \imath\Delta_{\pm\pm}(x;x^\prime)
\Big]\bigg\} \,. \label{Self-En-Big} \eea
The first line in~(\ref{Self-En-Big}) corresponds to the graviton
vacuum polarization tensor of a minimally coupled massless scalar
field already calculated in~\cite{Woodard:2010}. Our analysis in
Appendix~C shows that this expression can be also obtained by
expanding the 2PI bubble diagram around the Minkowski space, {\it
i.e.} take Eq.~(\ref{Sigma:general}) in the limit when
$g_{\mu\nu}\rightarrow\eta_{\mu\nu}$. The vacuum
polarization~(\ref{Self-En-Big}) contains a local divergence. In
order to extract it, the following identities can be
employed~\cite{Woodard:2005}, which are easily derived
from~(\ref{massless scalar propagators}--\ref{distance
functions}):
\bea
\partial'_\beta\partial_\alpha\imath\Delta_{++}(x;x^\prime)
&=&\delta_\alpha^0\delta_\beta^0 \imath\delta^D(x-x^\prime)
+\frac{\Gamma\left(\frac{D}{2}\right)}{2\pi^{D/2}}
\left[\frac{\eta_{\alpha\beta}}{(\Delta x_{++}^2)^{D/2}}
-D\frac{\Delta x_\alpha\Delta x_\beta}{(\Delta
x_{++}^2)^{(D/2)+1}}\right],
\nonumber\\
\partial'_\beta\partial_\alpha \imath\Delta_{--}(x;x^\prime)
&=&-\delta_\alpha^0\delta_\beta^0 i\delta^D(x-x^\prime)
+\frac{\Gamma\left(\frac{D}{2}\right)}{2\pi^{D/2}}
\left[\frac{\eta_{\alpha\beta}}{(\Delta x_{--}^2)^{D/2}}
-D\frac{\Delta x_\alpha\Delta x_\beta}{(\Delta
x_{--}^2)^{(D/2)+1}}\right],
\nonumber\\
\partial'_\beta\partial_\alpha \imath\Delta_{-+}(x;x^\prime)
&=&\frac{\Gamma\left(\frac{D}{2}\right)}{2\pi^{D/2}}
\left[\frac{\eta_{\alpha\beta}}{(\Delta x_{-+}^2)^{D/2}}
-D\frac{\Delta x_\alpha\Delta x_\beta}{(\Delta
x_{-+}^2)^{(D/2)+1}}\right],
\nonumber\\
\partial'_\beta\partial_\alpha\imath\Delta_{+-}(x;x^\prime)
&=&\frac{\Gamma\left(\frac{D}{2}\right)}{2\pi^{D/2}}
\left[\frac{\eta_{\alpha\beta}}{(\Delta x_{+-}^2)^{D/2}}
-D\frac{\Delta x_\alpha\Delta x_\beta}{(\Delta
x_{+-}^2)^{(D/2)+1}}\right] \,. \label{propagator:double
derivative} \eea
In order to reduce the $\xi$-dependent terms we also need the
following derivatives:
\bea
\partial_\sigma\partial^\prime_{\rho^\prime}\partial^\prime_{\sigma^\prime}
\imath\Delta_{\pm\pm}(x;x^\prime)
&=&\big(\delta^0_\sigma\delta_{\rho^\prime}^0\partial^\prime_{\sigma^\prime}
       + \delta^0_\sigma\delta_{\sigma^\prime}^0\partial^\prime_{\rho^\prime}
       - \delta^0_{\sigma^\prime}\delta_{\rho^\prime}^0\partial_\sigma\big)
    \imath\delta^D(x-x^\prime)
\nonumber\\
&+&D\frac{\Gamma\left(\frac{D}{2}\right)}{2\pi^{D/2}}
\bigg[\frac{\eta_{\rho^\prime\sigma^\prime}\Delta x_\sigma
   + \eta_{\rho^\prime\sigma}\Delta x_{\sigma^\prime}
   +\eta_{\sigma^\prime\sigma}\Delta x_{\rho^\prime}}
         {\Delta x_{\pm\pm}^{D+2}}
  -(D+2)\frac{\Delta x_{\rho^\prime}\Delta x_{\sigma^\prime}
            \Delta x_{\sigma}}{\Delta x_{\pm\pm}^{D+4}}
\bigg]
\nonumber\\
\partial_\rho\partial_\sigma\partial^\prime_{\rho^\prime}
       \partial^\prime_{\sigma^\prime}\imath\Delta_{\pm\pm}(x;x^\prime)
\!&=&\!\big(\delta^0_\sigma\delta_{\rho^\prime}^0
              \partial^\prime_{\sigma^\prime}\partial_{\rho}
  \!+\! \delta^0_{\sigma^\prime}\delta_{\rho}^0
              \partial_{\sigma}\partial^\prime_{\rho^\prime}
  \!+\! \delta^0_\sigma\delta_{\sigma^\prime}^0
              \partial^\prime_{\rho^\prime}\partial_{\rho}
  \!+\! \delta^0_{\rho^\prime}\delta_{\rho}^0
              \partial_{\sigma}\partial^\prime_{\sigma^\prime}
  \!-\! \delta^0_\sigma\delta_{\rho}^0
              \partial^\prime_{\sigma^\prime}\partial^\prime_{\rho^\prime}
  \!-\! \delta^0_{\sigma^\prime}\delta_{\rho^\prime}^0
              \partial_{\sigma}\partial_{\rho}
\big)
   \imath\delta^D(x\!-\!x^\prime)
\nonumber\\
&+&D\frac{\Gamma\left(\frac{D}{2}\right)}{2\pi^{D/2}}
\bigg[\frac{\eta_{\rho^\prime\sigma^\prime}\eta_{\rho\sigma}
 + \eta_{\rho^\prime\sigma}\eta_{\sigma^\prime\rho}
 +\eta_{\sigma^\prime\sigma}\eta_{\rho^\prime\rho}}{\Delta x_{\pm\pm}^{D+2}}
+(D+2)(D+4)\frac{\Delta x_\rho \Delta x_\sigma\Delta
x_{\rho^\prime}
        \Delta x_{\sigma^\prime}}{\Delta x_{\pm\pm}^{D+6}}
 \nonumber\\
&&\hskip -4cm -\,(D+2)\frac{\eta_{\rho^\prime\sigma^\prime}\Delta
x_{\rho}\Delta x_{\sigma} +\eta_{\rho^\prime\sigma}\Delta
x_{\rho}\Delta x_{\sigma^\prime} +\eta_{\rho^\prime\rho}\Delta
x_{\sigma^\prime}\Delta x_{\sigma} +\eta_{\rho\sigma^\prime}\Delta
x_{\rho^\prime}\Delta x_{\sigma} +\eta_{\rho\sigma}\Delta
x_{\rho^\prime}\Delta x_{\sigma^\prime}
+\eta_{\sigma\sigma^\prime}\Delta x_{\rho}\Delta x_{\rho^\prime}
               }
{\Delta x_{\pm\pm}^{D+4}}
\bigg] \,.\qquad \label{propagator:3+4 derivative} \eea
When these expressions are inserted into the graviton vacuum
polarization~(\ref{Self-En-Big}), the terms
in~(\ref{propagator:double derivative}--\ref{propagator:3+4
derivative}) that contain delta functions do not contribute in
dimensional regularization, since they hit a propagator which
contains a function of the type $(\Delta x)^{-\alpha}$, where
$\alpha=\alpha(D)$ is a D-dependent power. One can show that
rigorously by the analogous reasoning as was used in
Sec.~\ref{Local contributions}.

 The next step in the reduction of the graviton vacuum polarization tensor
is the extraction of derivatives. In the following we express
various terms appearing in the vacuum polarization as an operator
acting on the single function, $1/\Delta x^{2D-4}$ (which still
contains the local divergence that we are after):
%
\begin{align}
\frac{1}{\Delta x_{\pm\pm}^{2D}} &=\left\{
\frac{\partial^4}{4D(D-1) (D-2)^2} \right\}
  \frac{1}{\Delta x_{\pm\pm}^{2D-4}},
\\
\frac{\Delta x_\mu\Delta x_\nu}{\Delta x_{\pm\pm}^{2D+2}}
&=\left\{
\frac{\eta_{\mu\nu}\partial^4+D\partial_\mu\partial_\nu\partial^2}
         {8D^2(D-1) (D-2)^2} \right\}\frac{1}{\Delta
         x_{\pm\pm}^{2D-4}},
\\
\frac{\Delta x_\mu\Delta x_\nu\Delta x_\rho\Delta x_\sigma}
          {\Delta x_{\pm\pm}^{2D+4}}
&= \bigg\{\frac{\left(\eta_{\mu\nu}\eta_{\rho\sigma}
  +2\eta_{\mu(\rho}\eta_{\sigma)\nu}\right)\partial^4}{16D^2(D+1)(D-1)(D-2)^2},
\\ \nonumber
&\hskip 0.5cm
  +\,\frac{\left(\eta_{\mu\nu}\partial_\rho\partial_\sigma
  +4\partial_{(\mu}\eta_{\nu)(\rho}\partial_{\sigma)}
  +\eta_{\rho\sigma}\partial_\mu\partial_\nu\right)\partial^2
  +(D-2)\partial_\mu\partial_\nu\partial_\rho\partial_\sigma
         }{16D(D+1)(D-1)(D-2)^2}
%
\bigg\}\frac{1}{\Delta x_{\pm\pm}^{2D-4}} \,.
\end{align}
%
%
Before reducing all the terms of the tensor~(\ref{Self-En-Big}) it
is instructive to note the structure of the parts linear and
quadratic in $\xi$ which are basically the same up to a 
multiplicative constant. To be more precise, in the second and
third terms that are linear in $\xi$ the following operator can be
extracted:
\beq -(D^2-D-2)D_{\mu\nu}D_{\rho\sigma},
 \eeq
while in the term that is quadratic in $\xi$ we can extract
\beq 4(D^2-1)D_{\mu\nu}D_{\rho\sigma}. \eeq
After some algebra, and after combining the linear and quadratic
contributions in $\xi$, the graviton vacuum
polarization~(\ref{Self-En-Big}) can be written as
\bea
\imath\left[{}^{\pm}_{\mu\nu}\Pi_{\rho\sigma}^{\pm}\right](x;x^\prime)
&=&\left(\frac{\pm \kappa}{2}\right)\left(\frac{\pm
\kappa}{2}\right)
\frac{\Gamma^2\left(\frac{D}{2}\right)}{16\pi^D(D^2-1)(D-2)^2}
\\ \nonumber
&&\times\,\Bigg\{4D_{\alpha\beta\mu\nu}D^{\alpha\beta}{}_{\rho\sigma}
+\left[-\frac{D}{D-1}+8(D^2-1)\left(\xi-\frac{D-2}{4(D-1)}\right)^2\right]D_{\mu\nu}D_{\rho\sigma}\Bigg\}
\frac{1}{\Delta x_{\pm\pm}^{2D-4}} \,.
\label{Self-En-F1} \eea
Upon performing one more partial integration:
\beq \frac{1}{\Delta
x_{\pm\pm}^{2D-4}}=\frac{\partial^2}{2(D-3)(D-4)}
 \frac{1}{\Delta x_{\pm\pm}^{2D-6}}
\,, \label{xx} \eeq
we see that $(D-4)$ appears in the denominator, signifying a
divergence. This divergence can be removed by recalling the
identities (that were employed when constructing the
propagators~(\ref{massless scalar propagators})),
\bea
\partial^2\frac{1}{\Delta x_{++}^{D-2}}
 &=&\frac{ 4\pi^{D/2}}{\Gamma\left(\frac{D}{2}-1\right)}
  \imath\delta^D(x-x^\prime)
\,,\qquad \quad
\partial^2\frac{1}{\Delta x_{+-}^{D-2}}=0,
\\ \nonumber
\partial^2\frac{1}{\Delta x_{--}^{D-2}}
 &=&-\frac{ 4\pi^{D/2}}{\Gamma\left(\frac{D}{2}-1\right)}
         \imath\delta^D(x-x^\prime)
\,,\qquad\; \partial^2\frac{1}{\Delta x_{-+}^{D-2}}=0 \,.
\label{Delte} \eea
Based on these identities and Eq.~(\ref{xx}) we easily obtain
\bea \frac{1}{\Delta x_{\pm\pm}^{2D-4}}
 &=&\frac{\partial^2}{2(D-3)(D-4)}\left(\frac{1}{\Delta x_{\pm\pm}^{2D-6}}
    -\frac{\mu^{D-4}}{\Delta x_{\pm\pm}^{D-2}}\right)
 +(\sigma^3)^{\pm\pm}
    \frac{2\pi^{D/2}\mu^{D-4}}{\Gamma\big(\frac{D}{2}-1\big)(D-3)(D-4)}
   \imath\delta^D(x-x^\prime)
\,, \quad \label{++Expand} \eea
where $\sigma^3={\rm diag}(1,-1)$ is the Pauli matrix and $\mu$ is
the mass scale appearing for dimensional reasons which signifies
the renormalization scale. The identity~(\ref{++Expand}) is
judiciously constructed, such that $(\Delta x_{\pm\pm}^2)^{2-D}$
has been split into a piece that is finite in $D=4$ and a local
divergent piece. Indeed, when expanded around $D=4$,
Eq.~(\ref{++Expand}) can be recast as
\bea \frac{1}{\Delta x_{\pm\pm}^{2D-4}}
 &=& -\frac{\mu^{2D-8}\partial^2}{32}
       \left[\ln^2(\mu^2\Delta x_{\pm\pm}^2)-2\ln(\mu^2\Delta x_{\pm\pm}^2)
       +{\cal O}(D-4)\right]
\nonumber\\
  && +\,(\sigma^3)^{\pm\pm}\frac{2\pi^{D/2}\mu^{D-4}}
                              {\Gamma\big(\frac{D}{2}-1\big)(D-3)(D-4)}
       \imath\delta^D(x-x^\prime)
\,,\quad\; \label{++ExpandFinal} \eea
where we made use of
\bea \frac{1}{\Delta x_{\pm\pm}^{2D-6}}-\frac{\mu^{D-4}}{\Delta
x_{\pm\pm}^{D-2}}
&=& \frac{\mu^{2D-8}}{\Delta x_{\pm\pm}^2}
     \left[-\frac{D-4}{2} \ln{(\mu^2\Delta x_{\pm\pm}^2)}
          +\mathcal O \left((D-4)^2\right)
     \right]
\, \eea
and \beq
  \frac{\ln(\mu^2\Delta x_{\pm\pm}^2)}{\Delta x_{\pm\pm}^2}
      = \frac18\partial^2\Big[\ln^2(\mu^2\Delta x_{\pm\pm}^2)
                             -2\ln(\mu^2\Delta x_{\pm\pm}^2)\Big]
\,. \label{log/Delta x2} \eeq

With the results outlined in this Appendix it is possible now to
express a graviton polarization tensor in terms of the sum of the
non-local finite and local divergent
parts~(\ref{Sigma:split:ren+div}).

\end{document}